\documentclass[twocolumn,showpacs,preprintnumbers,amsmath,amssymb]{revtex4-1}
\usepackage{graphicx,epsfig}
\usepackage{dcolumn}
\usepackage{bm}
\usepackage{fixfoot}
\begin{document}

\title{Symmetry and localization of quantum walk induced by extra link in cycles}
\author{Xin-Ping Xu$^{1}$}
\author{Yusuke Ide$^{2}$}
\author{Norio Konno$^{3}$}
\affiliation{%
$^1$ School of Physical Science and Technology, Soochow University, Suzhou 215006, China\\
$^2$ Department of Information Systems Creation, Faculty of Engineering, Kanagawa University,
Yokohama, Kanagawa, 221-8686, Japan\\
$^3$ Department of Applied Mathematics, Faculty of Engineering, Yokohama National University, Hodogaya, Yokohama 240-8501, Japan
}%
\begin{abstract}
It is generally believed that the network structure has a profound impact on diverse dynamical processes
taking place on networks. Tiny change in the structure may cause completely different dynamics. In this paper,
we study the impact of single extra link on the coherent dynamics modeled by continuous-time quantum walks.
For this purpose, we consider the continuous-time quantum walk on the cycle with an additional link. We find that the additional link in cycle indeed cause a very different dynamical behavior compared to the dynamical behavior on the cycle. We analytically treat this problem and calculate the Laplacian spectrum for
the first time, and approximate the eigenvalues and eigenstates using the Chebyshev polynomial technique and perturbation theory. It is found that the probability evolution exhibits a similar behavior like the cycle if the exciton starts far away from the two ends of the added link. We explain this phenomenon by the eigenstate of the largest eigenvalue. We prove symmetry of the long-time averaged probabilities using the exact determinant equation for the eigenvalues expressed by Chebyshev polynomials. In addition, there is a significant localization when the exciton starts at one of the two ends of the extra link, we show that the localized probability is determined by the largest eigenvalue and there is a significant lower bound for it even in the limit of infinite system. Finally, we study the problem of trapping and show the survival probability also displays significant localization for some special values of network parameters, and we determine the conditions for the emergence of such localization. All our findings suggest that the different dynamics caused by the extra link in cycle is mainly determined by the largest eigenvalue and its corresponding eigenstate. We hope the Laplacian spectral analysis in this work provides a deeper understanding for the dynamics of quantum walks on networks.
\end{abstract}
 \maketitle
\section{Introduction}
The dynamic processes taking place in networks have attracted much attention in recent years~\cite{rn1,rn2,rn3}.
It is generally believed that network structure fundamentally influences the dynamical processes on networks. Investigation
on such aspect could be done using the spectral analysis and it has been shown that the dynamical behavior is related to the spectral properties of the networks~\cite{rn3,rn4}. Examples include synchronization of coupled dynamical systems~\cite{rn5}, epidemic spreading~\cite{rn6}, percolation~\cite{rn7}, community detection~\cite{rn8}, and others~\cite{rn3}. In several of these examples, the largest eigenvalue plays an important role in relevant dynamics.
All these examples suggest that spectral property is crucial to understanding the dynamical processes taking place in networks.

Quantum walks, as coherent dynamical process in network, have become a popular topic in the past few years~\cite{rn9,rn10,rn11,add0}. The continuous interest in quantum walk can be attributed to its broad applications to many distinct fields, such as polymer physics, solid state physics, biological
physics, and quantum computation~\cite{rn12,rn13}. In the literature~\cite{rn9,rn10}, there are two types of quantum walks: continuous-time and discrete-time quantum walks. It is shown that both type of quantum walks is closely related to the spectral properties of the Laplacian matrix of the network. Most of previous studies have studied quantum walks on some simple graphs, such as the line~\cite{rn14,add1}, cycle~\cite{rn15}, hypercube~\cite{rn16}, trees~\cite{rn17,add2}, dendrimers~\cite{rn18}, ultrametric spaces~\cite{add3}, threshold network~\cite{add4}, and other regular networks with simple topology~\cite{rn11,add0}. The quantum dynamics displays different behavior on different graphs, most of the conclusions hold solely in the particular geometry and how the structure influences the dynamics is still unknown. Because quantum walks have potential
applications in teleportation and cryptography in the field of quantum computation~\cite{rn13}, it is clearly beneficial to investigate how the structure influences the dynamics of quantum walks.

In this paper, we focus on continuous-time quantum walks (CTQWs) and study the impact of single extra link on its coherent dynamics. For this purpose, we consider the continuous-time quantum walk on the cycle with an additional link. The dynamics of continuous-time quantum walks on cycle is well known, the problem is analytically solvable and directly related to quantum carpets in solid state physics~\cite{rn15}. The topology of cycle is highly symmetric and the quantum dynamics reflects such topological symmetry. If one extra link is added into the cycle, the topological symmetry is broken and the structure has a small difference compared to the cycle. This enables us to treat the problem analytically and investigate how an additional link added to the network affects the dynamics. Similar study of the impact of single links on dynamical processes can be found in Refs.~\cite{rn19,rn20}, where the authors study the impact of single link addition in percolation~\cite{rn19} and impact of single link failure in quantum walks~\cite{rn20} respectively. In our case, in view of experimental implementation of CTQWs on cycles has been realized, we hope our study of the impact of single link (extra coupling) provides useful insight in quantum computation.

The rest of the paper is organized as follows. Section II introduces the model of continuous-time quantum walks and defines the network structure. Section III investigates the eigenvalues and eigenstates of the Laplacian matrix (Hamiltonian). We obtain the determinant equation for the eigenvalues expressed by Chebyshev polynomial, and calculate the largest eigenvalue on certain limit condition. The other eigenvalues and eigenstates are obtained by the perturbation theory. As we will show, like other dynamical processes, the largest eigenvalue and its eigenstate play an important role in the relevant dynamics.
In Section IV, we study the characteristics of probability evolution and distribution. The impact of the extra link depends on the initial exciton position, which can be well understood by the contribution of the eigenstate of the largest eigenvalue. The long time averaged probabilities shows significant symmetry and localization, and we give an explanation to these features using the determinant equation and largest eigenvalue. In Section V, we study the trapping process and the survival probabilities show significant localization for some special values of network parameters. We discuss this finding using the perturbation theory again. Conclusions and discussions are given in the last part, Sec. VI.

\section{Continuous-time quatum walks and the addition of link in cycle}
\subsection{Continuous-time quatum walks}
The coherent exciton transport on a connected network is modeled by
the continuous-time quantum walks (CTQWs), which is obtained by
replacing the classical transfer matrix by the Laplacian matrix , {\em i.e.}, $H=-T$~\cite{rn11,rn15}. The transfer matrix $T$
relates to the Laplacian matrix by $T=-\gamma A$, where for simplicity
we assume the transmission rates $\gamma$ of all bonds to be equal
and set $\gamma \equiv 1$ in the following~\cite{rn11,rn15}. The
Laplacian matrix $A$ has nondiagonal elements $A_{ij}$ equal to $-1$
if nodes $i$ and $j$ are connected and $0$ otherwise. The diagonal
elements $A_{ii}$ equal to degree of node $i$, {\em i.e.}, $A_{ii}=k_i$.
The states $|j\rangle$ endowed with the node $j$ of the network form
a complete, ortho-normalised basis set, which span the whole
accessible Hilbert space. The time evolution of a state $|j\rangle$
starting at time $t_0$ is given by $|j,t\rangle =
U(t,t_0)|j\rangle$, where $U(t,t_0)=exp[-iH(t-t_0)]$ is the quantum
mechanical time evolution operator. The transition amplitude
$\alpha_{k,j}(t)$ from state $|j\rangle$ at time $0$ to state
$|k\rangle$ at time $t$ reads $\alpha_{k,j}(t)=\langle
k|U(t,0)|j\rangle$ and obeys Schr\"{o}dinger¡¯s
equation~\cite{rn11,rn15}. The classical and quantum transition
probabilities to go from the state $|j\rangle$ at time $0$ to the
state $|k\rangle$ at time $t$ are given by $p_{k,j}(t)=\langle
k|e^{-tA}|j\rangle$ and $\pi_{k,j}(t)=|\alpha_{k,j}(t)|^2=|\langle
k|e^{-itH}|j\rangle|^2$~\cite{rn11,rn15}, respectively. Using $E_n$
and $|\Psi_n\rangle$ to represent the $n$th eigenvalue and
orthonormalized eigenstate of $H$, the quantum amplitudes
between two nodes can be written as~\cite{rn11}
\begin{equation}\label{eq1}
\alpha_{k,j}(t)=\sum_n e^{-itE_n}\langle k|\Psi_n\rangle \langle
\Psi_n|j\rangle,
\end{equation}
\begin{equation}\label{eq2}
\begin{array}{ll}
\pi_{k,j}(t)&=|\alpha_{k,j}(t)|^2=|\displaystyle\sum_n e^{-itE_n}\langle k|\Psi_n\rangle \langle \Psi_n|j\rangle|^2\\
&=\displaystyle\sum_{n,l} e^{-it(E_n-E_l)}\langle k|\Psi_n\rangle \langle
\Psi_n|j\rangle \langle j|\Psi_l\rangle \langle \Psi_l|k\rangle.
\end{array}
\end{equation}

For finite networks, $\pi_{k,j}(t)$ do not decay ad infinitum but at
some time fluctuates about a constant value. This value is
determined by the long time average of $\pi_{k,j}(t)$
\begin{equation}\label{eq3}
\begin{array}{ll}
\chi_{k,j}&=\displaystyle\lim_{T\rightarrow \infty}\frac{1}{T}\int_0^T
\pi_{k,j}(t)dt\\
&=\displaystyle\sum_{n,l}\langle k|\Psi_n\rangle \langle \Psi_n|j\rangle \langle j|\Psi_l\rangle \langle \Psi_l|k\rangle \\
&\  \ \  \ \times\displaystyle\lim_{T\rightarrow \infty}\frac{1}{T}\int_0^T e^{-it(E_n-E_l)}dt\\
&=\displaystyle\sum_{n,l}\delta(E_n,E_l)\langle k|\Psi_n\rangle \langle
\Psi_n|j\rangle \langle j|\Psi_l\rangle \langle \Psi_l|k\rangle.
\end{array}
\end{equation}
where $\delta(E_n,E_l)$ takes value 1 if $E_n$ equals to $E_l$ and
0 otherwise. To calculate the exact analytical expressions for $\pi_{k,j}(t)$
and $\chi_{k,j}$, all the eigenvalues $E_n$ and eigenstates
$|\Psi_n\rangle$ of the Laplacian matrix are required. If all the eigenvalues
$\{E_i\ |i=1,2,...,N\}$ are distinct, {\em i.e.}, all the eigenvalues are not degenerated, Eq.~(\ref{eq3}) can be simplified as,
\begin{equation}\label{eq4}
\chi_{k,j}=\sum_{n} |\langle k|\Psi_n\rangle|^2 \cdot |\langle
\Psi_n|j\rangle|^2.
\end{equation}
The eigenvalues for the cycle graph is twofold degenerated. However, as we will show, the addition of single link causes
the degeneracy to disappear. Therefore we can use Eq.~(\ref{eq4}) to calculate the probability distribution for our model. The transition prababilities (See Eqs.~(\ref{eq2})-(\ref{eq4})) are closely related to $E_n$ and $|\Psi_n\rangle$, it is crucial to calculate the Laplacian eigenspectrum and we will do this in Sec. III.
\begin{figure}
\scalebox{0.3}[0.3]{\includegraphics{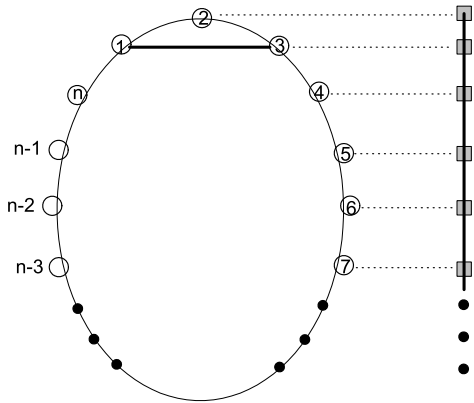}} \caption{Topology of the cycle with an extra link.
The additional link connects node $1$ and $m\equiv3$ in the cycle. According to the symmetry of topology, the graph can be projected into 1D chain
of size $N'\approx \lceil N/2\rceil$ (the notation $\lceil N/2\rceil$ denotes the integer part of $N/2$).
 \label{fg1}}
\end{figure}
\subsection{Addition of link in cycle}
Now we specify the topology of the cycle with an extra link. First we construct a cycle of size $N$ where each node connected to its two nearest-neighbor
nodes, then we connect two node of certain distance with an additional link. The topology is completely determined by the distance and size of the cycle.
For the sake of simplicity, we assign a consecutive number from $1$ to $N$ for each neighbored node in the cycle. The additional link connects node $1$ and
node $m$ in the cycle. Thus the structure, denoted by $G(N,m)$, are characterized by the network size $N$ and connecting node $m$ ($m\in[3,N-1]$). Here, we use $d(1,m)$ to denote the (shortest) distance between node $1$ and node $m$ in the cycle (without an extra link), the graph can also be characterized by $G(N,d(1,m))$. The structure of $G(N,m=3)$ is illustrated in Fig.~\ref{fg1}.

It is interesting to note that the network considered also has a symmetric structure, and the symmetry axis lies at the central position between node $1$ and $m$. The structure can be mapped into a 1D chain of size $N'\approx \lceil N/2\rceil$ (the notation $\lceil N/2\rceil$ denotes the integer part of $N/2$) if we make a horizontal projection of the graph (See Fig.~\ref{fg1}). Such projection sheds some light on the implicit relationship between the two structures. For example, as we show in Appendix C, the eigenvalues of the Laplacian matrix of the projected 1D chain also belong to eigenvalues of our model.

\section{Laplacian Eigenvalues and Eigenstates}
Since the transition probabilities are determined by the Laplacian eigenvalues and eigenstates (see Eqs.~(\ref{eq2})-(\ref{eq4})), a detailed analysis to the
eigenvalues and eigenstates will be helpful for the problem. In this section, we will study the Laplacian eigenvalues and eigenstates in detail.

\begin{figure*}
\scalebox{0.4}[0.4]{\includegraphics{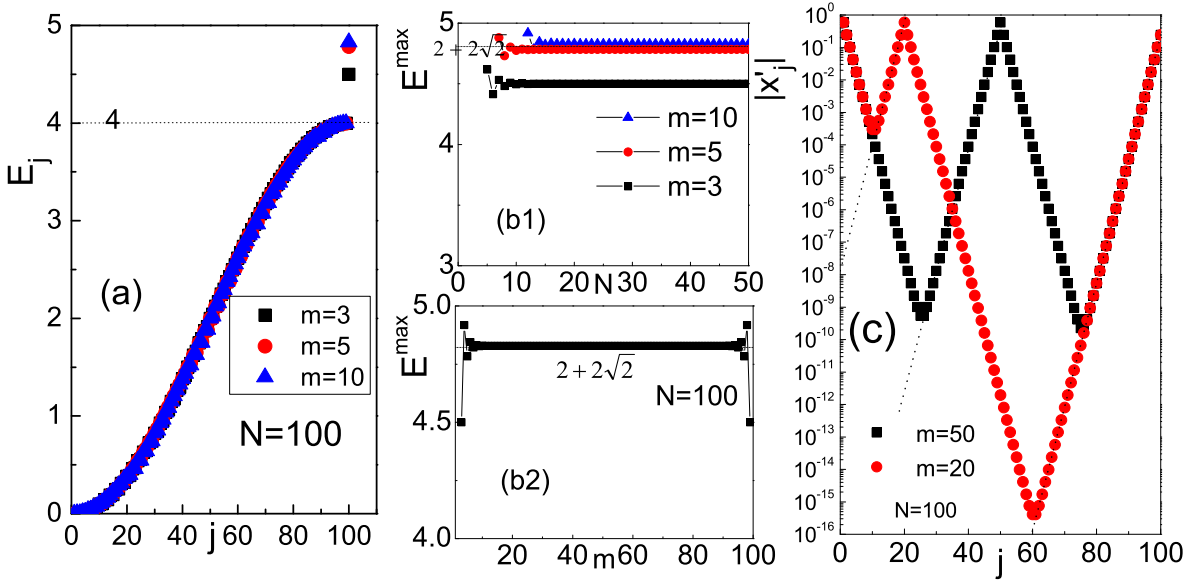}} \caption{(Color online) (a)Eigenvalues $E_j$, arranged in ascending
order, for the networks of $N=100$ with $m=3$, $m=5$ and $m=10$. The largest eigenvalue is larger than 4 and isolated from the other
eigenvalues. (b1) shows the largest eigenvalue $E^{max}$ as a function of system size $N$ for $m=3$, $m=5$ and $m=10$. (b2) shows the largest eigenvalue $E^{max}$ as a function of $m$ for fixed network size $N=100$. In large system and long range coupling, {\em i.e.}, the distance between node $m$ and node $1$ ($d(1,m)$) is not small (the plat region in the plot), the largest eigenvalue approaches to constant value $(2+2\sqrt{2})$. (c) Components of eigenstate of the largest eigenvalue $|x'_j|$ as a function of $j$ for networks of size $N=100$ with $m=20$ (red dots) and $m=50$ (black squares). The decay is exponential and there are two maximal points ($|x'_1|$ and $|x'_m|$), the distribution is symmetric, $|x'_j|=|x'_{m+1-j}|,\  \forall \ j\in [1,m]$, $|x'_j|=|x'_{m+{\scriptscriptstyle N}+1-j}|,\ \forall \ j\in [m+1,N]$.\label{fg2}}
\end{figure*}

Figure~\ref{fg2}(a) shows the eigenvalues obtained by numerical diagonalizing the Laplacian matrix using the software Mathematica 7.0.
The eigenvalues are ranked in ascending order for networks of size $N=100$ with $m=3$, $m=5$ and $m=10$. It is interesting to note that the largest eigenvalue
is larger than $4$ and isolated from the other eigenvalues (the other eigenvalues are less than $4$), this means the gap between the largest eigenvalue and the second largest eigenvalue does not converge to zero when the size of the system goes to infinity, whereas the other nearest eigenvalue gaps tend to zero in the limit of infinite system. This suggests the eigenvalue spectra is continuous except the largest eigenvalue. Such characteristic feature of eigenvalue spectra is useful for understanding the coherent dynamical behavior. As we will show, the
largest eigenvalue plays an important role in the dynamics and gives significant contribution to the localized probabilities. Hence in the following, we will try to obtain Laplacian eigenspectrum and determine the largest eigenvalue and its corresponding eigenstate.

\subsection{Determinant equation for the eigenvalues}
We start our analysis on the eigen equation of the Hamilton(Laplacian matrix). The Laplacian matrix $H$ of $G(N,m)$ ($m \geqslant 3$) takes the following form:
\begin{equation}\label{eq5}
 H_{ij}=\langle i|H|j\rangle=\left\{
\begin{array}{ll}
3,   & {\rm if} \ i=j=1,\ \ {\rm or} \ \ i=j=m\\
2,   & {\rm if} \ i=j\neq 1,\ \ {\rm and} \ \ i=j\neq m\\
-1,   & {\rm if} \ i \ {\rm and} \ j \ {\rm connected}  \\
 0,    & {\rm otherwise}.
\end{array}
\right.
\end{equation}
According to the eigen equation $H|\Psi\rangle=E|\Psi\rangle$, suppose the eigenstate $|\Psi\rangle$ can be expressed as
\begin{equation}\label{eq6}
|\Psi\rangle=\sum_{i=1}^Nx_i|i\rangle,
\end{equation}
The eigen equation can be decomposed into the following $N$ linear equations,
\begin{align}
3x_1-x_2-x_m-x_{\scriptscriptstyle N}&=Ex_1, \label{eq7} \\
-x_{j-1}+2x_j-x_{j+1}&=Ex_j, \ \ \ 1<j<m    \label{eq8} \\
-x_1-x_{m-1}+3x_m-x_{m+1}&=Ex_m,      \label{eq9}  \\
-x_{j-1}+2x_j-x_{j+1}&=Ex_j, \ \ \ \label{eq10}  m<j<N \\
-x_{{\scriptscriptstyle N}-1}+2x_{\scriptscriptstyle N}-x_1&=Ex_{\scriptscriptstyle N}. \ \ \ \label{eq11}
\end{align}
Eq.~(\ref{eq8}) can be rewritten as $(2-E)x_j=x_{j-1}+x_{j+1}$. This is similar to the recursive definition of the Chebyshev polynomials of the second kind (See Appendix A). Noting the recursive relations and setting $(2-E)\equiv 2x$ in the definition of Chebyshev polynomials, the variables $x_1, x_2,...,x_{m-2}$ in Eq.~(\ref{eq8}) can be expressed as a function of $x_{m-1}$ and $x_m$,
\begin{equation}\label{eq12}
x_{j-1}=U_{m-j}(x)x_{m-1}-U_{m-1-j}(x)x_m. \ \ 1<j\leqslant m
\end{equation}
Analogously, Eqs.~(\ref{eq10}) and (\ref{eq11}) can be written as a function of $x_{\scriptscriptstyle N}$ and $x_1$ using the Chebyshev polynomials,
\begin{equation}\label{eq13}
x_{j-1}=U_{{\scriptscriptstyle N}+1-j}(x)x_{\scriptscriptstyle N}-U_{{\scriptscriptstyle N}-j}(x)x_1.  \ \ m<j\leqslant N
\end{equation}
Substitute $E=2-2x$ into Eqs.~(\ref{eq7}) and (\ref{eq9}), we obtain,
\begin{align}
3x_1-x_2-x_m-x_{\scriptscriptstyle N}&=(2-2x)x_1, \label{eq14} \\
-x_1-x_{m-1}+3x_m-x_{m+1}&=(2-2x)x_m. \label{eq15}
\end{align}
In Eq.~(\ref{eq12}), $x_{1}=U_{m-2}(x)x_{m-1}-U_{m-3}(x)x_m$, $x_{2}=U_{m-3}(x)x_{m-1}-U_{m-4}(x)x_m$.  Replacing the variables $x_1$ and $x_2$ in
Eqs.~(\ref{eq14}) and (\ref{eq15}), we get,
\begin{equation} \label{eq16}
\begin{aligned}
x_{\scriptscriptstyle N}&=[(2x+1)U_{m-2}(x)-U_{m-3}(x)]x_{m-1}+ \\
& \ \ \ \  [U_{m-4}(x)-(2x+1)U_{m-3}(x)-1]x_m,
\end{aligned}
\end{equation}
\begin{equation} \label{eq17}
\begin{aligned}
x_{m+1}&=-(1+U_{m-2}(x))x_{m-1}+ \\
&\ \ \ \ (2x+1+U_{m-3}(x))x_m.
\end{aligned}
\end{equation}
Substituting $x_{1}=U_{m-2}(x)x_{m-1}-U_{m-3}(x)x_m$ and $x_{\scriptscriptstyle N}$ in Eq.~(\ref{eq16}) into Eq.~(\ref{eq13}) for $j=m+1$ and $j=m+2$, we obtain,
\begin{equation}\label{eq18}
c_1x_{m-1}+c_2x_m=0,
\end{equation}
where $c_1=U_{{\scriptscriptstyle N}-m}(x)[(2x+1)U_{m-2}(x)-U_{m-3}(x)]-U_{{\scriptscriptstyle N}-m-1}(x)U_{m-2}(x)$ and $c_2=U_{{\scriptscriptstyle N}-m}(x)[U_{m-4}(x)-(2x+1)U_{m-3}(x)-1]+U_{{\scriptscriptstyle N}-m-1}(x)U_{m-3}(x)-1$, and
\begin{equation} \label{eq19}
\begin{aligned}
x_{m+1}&=\{U_{{\scriptscriptstyle N}-m-1}(x)[(2x+1)U_{m-2}(x)-U_{m-3}(x)] \\
&\ \ \ -U_{{\scriptscriptstyle N}-m-2}(x)U_{m-2}(x)\}x_{m-1}+ \\
&\ \ \ \{U_{{\scriptscriptstyle N}-m-1}(x)[U_{m-4}(x)-(2x+1)U_{m-3}(x)-1] \\
&\ \ \ +U_{{\scriptscriptstyle N}-m-2}(x)U_{m-3}(x)\}x_m.
\end{aligned}
\end{equation}
Combine Eq.~(\ref{eq17}) and Eq.~(\ref{eq19}), we get another equation for $x_{m-1}$ and $x_m$,
\begin{equation}\label{eq20}
c_3x_{m-1}+c_4x_m=0,
\end{equation}
where $c_3=U_{{\scriptscriptstyle N}-m-1}(x)[(2x+1)U_{m-2}(x)-U_{m-3}(x)]-U_{{\scriptscriptstyle N}-m-2}(x)U_{m-2}(x)+U_{m-2}(x)+1$ and $c_4=U_{{\scriptscriptstyle N}-m-1}(x)[U_{m-4}(x)-(2x+1)U_{m-3}(x)-1] +U_{{\scriptscriptstyle N}-m-2}(x)U_{m-3}(x)-U_{m-3}(x)-(2x+1)$.
Thus we have got two equations for $x_{m-1}$ and $x_m$, Eq.~(\ref{eq18}) and Eq.~(\ref{eq20}). The two equations should have nonzero solutions, which leads to,
\begin{equation}\label{eq21}
c_1c_4-c_2c_3=0.
\end{equation}
In the Appendix B, we show the substraction of the products can be simplified to be a much simple form in Eq.~(\ref{b7}),
thus we have obtained determinant equation for the eigenvalues,
\begin{equation}\label{eq22}
1+U_{{\scriptscriptstyle N}-m}(x)+U_{m-2}(x)-U_{{\scriptscriptstyle N}-1}(x)-T_{\scriptscriptstyle N}(x)=0,
\end{equation}
where $T_{\scriptscriptstyle N}(x)$ and $U_{{\scriptscriptstyle N}-1}(x)$ are Chebyshev polynomials of the first kind and the second kind, respectively.
Solving the above equation, we can get all the Laplacian eigenvalues. This equation is useful to determine the largest eigenvalue and interpret the
symmetric structure of the transition probabilities.

\subsection{The largest eigenvalue}
The largest eigenvalue can be determined using Eq.~(\ref{eq22}) under certain limit conditions. As we have shown, the largest eigenvalue is larger than $4$, this corresponds to the solution $x<-1$ in Eq.~(\ref{eq22}). Note that the Chebyshev polynomials is divergent for $|x|>1$ in the limit of infinite order, Eq.~(\ref{eq22}) divided by $T_N(x)$ leads to,
\begin{equation}\label{eq23}
\frac{U_{{\scriptscriptstyle N}-m}(x)}{T_{\scriptscriptstyle N}(x)}+\frac{U_{m-2}(x)}{T_{\scriptscriptstyle N}(x)}-\frac{U_{{\scriptscriptstyle N}-1}(x)}{T_{\scriptscriptstyle N}(x)}-1=0.
\end{equation}
For large size of system and long range coupling, {\em i.e.}, $N$ is large and the distance $d(1,m)$ between node $1$ and node $m$ is not small, if we apply the asymptotic solution of the Chebyshev polynomials (See Eq.~(\ref{a5}) in Appendix A), the first two terms of the above equation equal to $0$, $-\frac{U_{{\scriptscriptstyle N}-1}(x_0)}{T_{\scriptscriptstyle N}(x_0)}\approx \frac{2}{|z_0^{-1}-z_0|}=1$, which leads to $z_0=-1-\sqrt{2}$ (the other solutions do not satisfy $z=x-\sqrt{x^2-1}$ less than $-1$ when $x<-1$) and $x_0=\frac{1+z_0^2}{2z_0}=-\sqrt{2}$. Thus the largest eigenvalue equals to a constant value $E^{max}=2-2x_0=2+2\sqrt{2}$ under this limit condition.

To test the above prediction, we plot the largest eigenvalue as a function of $N$ and $m$ in Fig.~\ref{fg2}(b). As we can see, the largest eigenvalue converges
to a constant value as $N$ or $m$ increases. For moderate range coupling, {\em e.g.}, $m=5$, the largest eigenvalue is close to the analytical prediction $(2+2\sqrt{2})$. The constant value of the largest eigenvalue suggests that its corresponding eigenstate (eigenvector) approaches to certain stationary distribution. Here, the obtained largest eigenvalue is useful to consider its corresponding eigenstate, which
we will show in the following.
\subsection{Eigenstate of the largest eigenvalue}
The components of eigenstate corresponding to the largest eigenvalue are shown in Fig.~\ref{fg2}(c). The straight line in the linear-log plot indicates that
the components display an exponential decay, $i.e.$, $|x'_j|\sim z_0^{-j}$. The distribution of the components of eigenstate is symmetric, $|x'_j|=|x'_{m+1-j}|,\  \forall \ j\in [1,m]$; $|x'_j|=|x'_{m+{\scriptscriptstyle N}+1-j}|,\ \forall \ j\in [m+1,N]$. As we will prove in Sec. IV (subsection B) and Appendix C, this symmetric behavior is true for all the eigenstates.

To obtain an approximate formula for the eigenstate of the largest eigenvalue, we use Eq.~(\ref{eq18}) to present $x_{m-1}$ as a function of $x_m$, and substitute it into Eq.~(\ref{eq16}) to write $x_{\scriptscriptstyle N}$ as a function of $x_m$. Using the Chebyshev polynomial identity (\ref{a7}) in Appendix A, Eq.~(\ref{eq16}) can be written as $x_{\scriptscriptstyle N}=[U_{m-1}(x)+U_{m-2}(x)]x_{m-1}-[U_{m-2}(x)+U_{m-3}(x)+1]x_m$. According to Eq.~(\ref{eq18}), $x_{m-1}=-c_2x_m/c_1$ and noting the simplified $c_1$, $c_2$ in (\ref{b1}) and (\ref{b2}) in Appendix A, we obtain $x_{\scriptscriptstyle N}=\frac{c'_2}{c_1}x_m$ (See the detailed derivation in Eq.~(\ref{b8}) in Appendix B). Substituting $x_{m-1}$ and $x_{\scriptscriptstyle N}$ into Eqs.~(\ref{eq12}) and (\ref{eq13}) and noting $x_1=-x_m$ for the largest eigenvalue (See Eq.~(\ref{c12}) in Appendix C), Eqs.~(\ref{eq12}) and (\ref{eq13}) can be written as,
\begin{equation} \label{eq24}
\begin{aligned}
x_{j}&=-[\frac{c_2}{c_1}U_{m-j-1}(x)+U_{m-2-j}(x)]x_m, j\in[1,m] \\
x_{j}&=[\frac{c'_2}{c_1}U_{{\scriptscriptstyle N}-j}(x)+U_{{\scriptscriptstyle N}-j-1}(x)] x_m. \ \ \ j\in[m,N]
\end{aligned}
\end{equation}
For the largest eigenvalue, $x=x_0=-\sqrt{2}$, the ratios $c_2/c_1$ and $c'_2/c_1$ approach to the constant value $-z_0^{-1}=(\sqrt{2}-1)$ for large system and long range coupling. Therefore, Eq.~(\ref{eq24}) can be recasted as,
\begin{equation} \label{eq25}
\begin{aligned}
x'_{j}&\approx[ z_0^{-1}\cdot U_{m-1-j}(x_0)-U_{m-2-j}(x_0)  ] x'_m \\
&= z_0^{j-m}x'_m, \ \ j\in[\lceil\frac{m+1}{2}\rceil,m] \\
x'_{j}&\approx[  -z_0^{-1}\cdot U_{{\scriptscriptstyle N}-j}(x_0)+U_{{\scriptscriptstyle N}-1-j}(x_0)  ] x'_m  \\
&=-z_0^{j-N-1}x'_m. \ \ j\in[\lceil\frac{(m+N+1)}{2}\rceil,N]\\
\end{aligned}
\end{equation}

The above formula only describes the right part of the symmetric distribution of the eigenstate (See the points on the dashed lines in Fig.~\ref{fg2}(c)). The decay exponents in Eq.~(\ref{eq25}) only depend on the (shortest) distance between the node pairs $(j,1)$ and $(j,m)$, {\em i.e.}, $d_j=\text{min}\{d(j,1),d(j,m)\}$. Eq.~(\ref{eq25}) can be summarized in a much simple form using the shortest distance $d_j$,
\begin{equation} \label{eq26}
x'_{j}\approx \pm z_0^{-d_j}x'_m, \ \ d_j=\text{min}\{d(j,1),d(j,m)\}
\end{equation}
In the calculation, the eigenstate should be normalized, $\sum_{j=1}^N|x'_j|^2=1$. For large system and long-range couple, the summation has a week dependence on $N$ and $m$, thus the summation of finite geometric series approaches to the summation of the infinite geometric series, {\em i.e.},
\begin{equation} \label{eq27}
\begin{aligned}
\sum_{j=1}^N|x'_j|^2=&\sum_{j=1}^m|x'_j|^2+\sum_{j=m+1}^N|x'_j|^2 \\
\approx &\big(2\sum_{j=1}^{\infty}z_0^{2(1-j)} + 2\sum_{j=1}^{\infty}z_0^{-2j}\big)|x'_m|^2 \\
=& 2\sqrt{2}|x'_m|^2=1,
\end{aligned}
\end{equation}
which leads to $|x'_1|=|x'_m|=2^{-3/4}$. For large systems and long range coupling, the distribution of the eigenstate of the largest eigenvalue approaches to a stationary exponential decay with the maximal component $|x'_1|=|x'_m|=2^{-3/4}$. In Fig.~\ref{fg2}(c), we plot the analytical predictions in Eq.~(\ref{eq26}), which agrees well with the exact numerical result.

We would like to point out that the exponential distribution of the components of the eigenstate is an interesting characteristic for the largest eigenvalue.
For the other eigenstates, the components do not display this feature. This could be interpreted by the striking different behavior of
the Chebyshev polynomials for $|x|>1$ and $|x|<1$. For $|x|\leqslant 1$, the Chebyshev polynomials show finite regular oscillations, but for $|x|> 1$, Chebyshev polynomials show exponential growth. The largest eigenvalue and its eigenstate play a significant role in the coherent dynamics, we will discuss this in the next section.
\subsection{Other eigenvalues and eigenstates}
We have determined the largest eigenvalue and its corresponding eigenstate using the determinant equation (\ref{eq22}). However, it is not intuitionistic to determine the other eigenvalues and eigenstates by Eq.~(\ref{eq22}). Here, we use the perturbation theory to get the approximate results for the other eigenvalues and eigenstates.

According to perturbation theory, the Hamilton (Laplacian matrix) of the cycle with an additional link $H$ can be divided into two parts: the unperturbed Hamilton $H^{(0)}$ and perturbed Hamilton $H'$ ($H=H^{(0)}+H'$, $H|\Psi\rangle=E|\Psi\rangle$). The unperturbed Hamilton $H^{(0)}$ is the Laplacian matrix for the cycle, whose eigenvalues and eigenstates are well known: $H^{(0)}|\psi_n^{(0)}\rangle=E_n^{(0)}|\psi_n^{(0)}\rangle$, $E_n^{(0)}=2-2\cos\theta_n$ , $|\psi_n^{(0)}\rangle=\frac{1}{\sqrt{N}}\sum_{j=1}^Ne^{-ij\theta_n}|j\rangle$ ($n\in[-\lceil\frac{N}{2}\rceil,\lceil\frac{N}{2}\rceil]$, $\theta_n=2n\pi/N$). The perturbed Hamilton $H'$ can be written as $H'=|1\rangle\langle1|+|m\rangle\langle m|-|1\rangle\langle m|-|m\rangle\langle 1|$. The eigenvalues of the cycle $E_n^{(0)}$ are twofold degenerated except the minimal eigenvalue $0$ and maximal eigenvalue $4$. The first order corrections for eigenvalues $0$ and $4$ are $E_{0}^{(1)}=0$ and $E_{N/2}^{(1)}=2/N[1+(-1)^m]$ respectively (maximal eigenvalue $4$ only exists for even $N$). The zero order approximation of the eigenstates corresponding to $0$ and $4$ are equal to the original unperturbed eigenstates of $H^{(0)}$, {\em i.e.}, $|\Psi_0^{(0)}\rangle=|\psi_0^{(0)}\rangle=1/\sqrt{N}\sum_{j=1}^N|j\rangle$, $|\Psi_{N/2}^{(0)}\rangle=|\psi_{N/2}^{(0)}\rangle=1/\sqrt{N}\sum_{j=1}^N(-1)^j|j\rangle$. For the twofold degenerated states, the degeneracy disappears when an extra link is added. To approximate these eigenvalues and eigenstates, we apply degenerate perturbation theory to calculate the first order corrections of the eigenvalues and zero order approximation of the eigenstates.

We diagonalize the the perturbed Hamilton $H'$ using the unperturbed degenerated eigenstates $|\psi_n^{(0)}\rangle$ and $|\psi_{-n}^{(0)}\rangle$.
The perturbed Hamilton $H'$ can be written as,
\begin{equation}\label{eq28}
\begin{array}{ll}
H'=
\begin{pmatrix}
 H'_{n,n} & H'_{n,-n} \\
 H'_{-n,n} & H'_{-n,-n}
\end{pmatrix}, n\in[1,\lceil\frac{N}{2}\rceil-\lambda], \\
 \text{where} \ \
\lambda=\begin{cases}
1, & \text{if }N\text{ is even} \\
0, & \text{if }N\text{ is odd}
\end{cases}
\end{array}
\end{equation}
where $H'_{n,n}=H^{'*}_{-n,-n}=\langle \psi_n^{(0)}|H'|\psi_n^{(0)}\rangle=\frac{2}{N}[1-\cos(m-1)\theta_n]$, $H'_{n,-n}=H^{'*}_{-n,n}=\langle \psi_{n}^{(0)}|H'|\psi_{-n}^{(0)}\rangle=\frac{1}{N}[e^{i\theta_n}-e^{im\theta_n}]$. The first order corrections of the eigenvalues are given by the two eigenvalues in Eq.~(\ref{eq28}),
\begin{equation}\label{eq29}
E_n^{(1)}=0, \ \ \ E_{-n}^{(1)}=\frac{4}{N}[1-\cos(m-1)\theta_n]
\end{equation}
The corresponding eigenstates for $E_n^{(1)}$ and $E_{-n}^{(1)}$ are,
\begin{equation}\label{eq30}
K_{n}=\frac{1}{\sqrt{2}}\binom{e^{i(m+1)\theta_n}}{1},  K_{-n}=\frac{1}{\sqrt{2}}\binom{-e^{i(m+1)\theta_n}}{1}
\end{equation}
The zero order approximation sof the eigenstates are given by
\begin{equation}\label{eq31}
\begin{aligned}
|\Psi_n^{(0)}\rangle=&K_n(1)|\psi_n^{(0)}\rangle+K_n(2)|\psi_{-n}^{(0)}\rangle, \\
|\Psi_{-n}^{(0)}\rangle=&K_{-n}(1)|\psi_n^{(0)}\rangle+K_{-n}(2)|\psi_{-n}^{(0)}\rangle
\end{aligned}
\end{equation}
Substitute the Bloch states $|\psi_n^{(0)}\rangle=\frac{1}{\sqrt{N}}\sum_{j=1}^Ne^{-ij\theta_n}|j\rangle$ into the above equation, we obtain,
\begin{equation}\label{eq32}
\begin{aligned}
|\Psi_n^{(0)}\rangle=&\frac{1}{\sqrt{2N}}\sum_{j=1}^N(e^{ij\theta_n}+e^{-i(j-m-1)\theta_n})|j\rangle, \\
|\Psi_{-n}^{(0)}\rangle=&\frac{1}{\sqrt{2N}}\sum_{j=1}^N(e^{ij\theta_n}-e^{-i(j-m-1)\theta_n})|j\rangle
\end{aligned}
\end{equation}
Thus we have obtained the first order corrections of the degenerated eigenvalues and the zero order approximation of the corresponding eigenstates(See Eqs.~(\ref{eq29}) and (\ref{eq32})). Using these eigenvalues and eigenstates, we can calculate the transition probabilities in Eqs.~(\ref{eq2})-(\ref{eq4}).

It is worth mentioning that the perturbation theory can not be used to calculate the largest eigenvalue, since the eigenvalue gaps do
not converge to zero in this case. However, the analytical predicted eigenvalues by perturbation theory (See Eq.~(\ref{eq29})) can also be obtained by Taylor series expansion of the Chebyshev polynomials. In Appendix D, we approximate the other eigenvalues using the determinant equation (\ref{eq22}), which are consistent with the predictions in Eq.~(\ref{eq29}).
\section{Probability evolution and distribution}
In this section, we consider the probability evolution and probability distribution. We will explain the observed phenomenon using the obtained eigenvalues and eigenstates.
\begin{figure}
\scalebox{0.45}[0.45]{\includegraphics{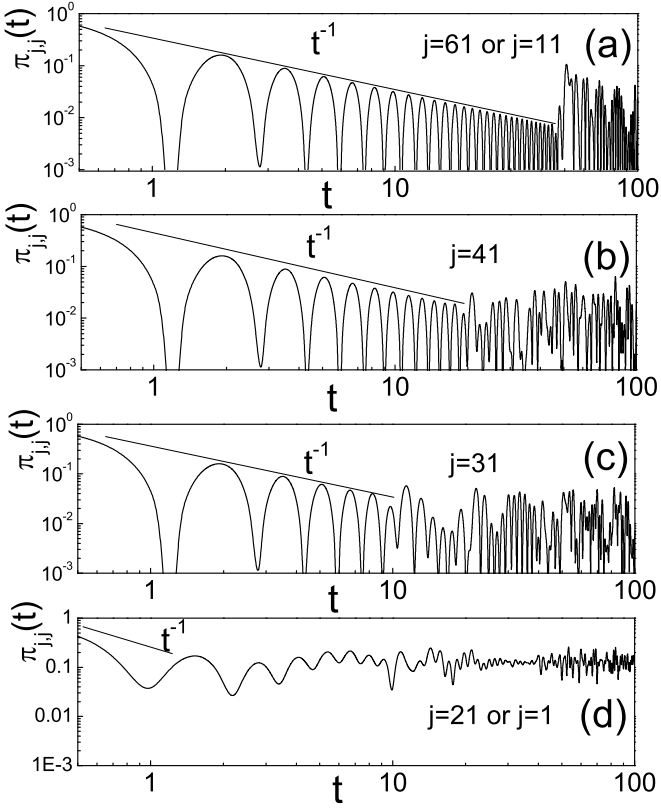}} \caption{Return probabilities $\pi_{j,j}(t)$ on a netwok of $N=100$ and $m=21$ for different starting positions $j$. (a)The exciton starts at the center node (symmetry axis) between $1$ and $m$, {\em i.e.}, $j=(m+1)/2=11$ or $j=(N+m+1)/2=61$. (b)Starting at $j=41$. (c)Starting at $j=31$. (d)Starting at $j=1$ or $j=m=21$. The time range of scaling $t^{-1}$ becomes short when the starting position near the two ends of the added link.
\label{fg3}}
\end{figure}
\subsection{Probability Evolution}
The time evolution of the probability is given by Eq.~(\ref{eq2}). Here, we focus on the return probability $\pi_{j,j}(t)$ (set $k=j$ in Eq.~(\ref{eq2})).
The decay behavior is used to quantify the efficiency of the transport~\cite{rn11}. In our case the return probability $\pi_{j,j}(t)$ depends on the starting position $j$.
We study the time evolution of the return probability $\pi_{j,j}(t)$ on different starting position $j$. The behavior of $\pi_{j,j}(t)$ versus $t$ are plotted in Fig.~\ref{fg3}. It is found that $\pi_{j,j}(t)$ shows different time range for scaling behavior $\pi_{j,j}(t)\sim t^{-1}$: For the starting positions far from the connected nodes $1$ and $m$, the scaling range of time is much larger than that of the walk starting near the nodes $1$ and $m$ (Compare scaling in the four figures). It is well known that for the cycle, $\pi_{j,j}(t)=J_0^2(2t)\sim t^{-1}$. This suggests that the additional link gives small influence on the dynamics when the walk starts at outlying positions, and gives larger influence for exciton near nodes $1$ and $m$. If the walk starts at the center nodes (nodes at symmetry axis) $j=(m+1)/2$ ($m$ is odd) or $j=(N+m+1)/2$ ($N+m$ is odd) (See Fig.~\ref{fg3} (a)), the probabilities are exact the same as the probabilities in cycle of the same size, in this case the additional link has no impact on the dynamics.

The observed phenomenon can be well understood using the eigenstates obtained in the previous section. If the exciton starts at positions far away from $1$ and $m$, the components of the eigenstate of the largest eigenvalue are very small due to the exponential decay of the components. Thus the contribution from the largest eigenvalue in Eq.~(\ref{eq1}) is very small, the main contribution to the amplitude comes from the other eigenvalues, which leads to the scaling behavior resembling to the case in cycles. If the exciton starts at the center nodes (nodes at symmetry axis) $j=(m+1)/2$ ($m$ is odd) or $j=(N+m+1)/2$ ($N+m$ is odd), the corresponding components of the largest eigenvalue's eigenstate are $0$ (See Eqs. (\ref{c13})-(\ref{c14}) in Appendix C). In this particular case, the largest eigenvalue gives no contribution to the amplitude, the probabilities are exact the same as in the cycles, thus the additional link has no impact on the dynamics. If the exciton starts at the positions near the connected nodes $1$ and $m$, the components of the largest eigenvalue's eigenstate become larger and give contribution to the amplitude, thus the scaling behavior disappears gradually (See Fig.~\ref{fg3}(b) (c)). If the exciton starts at nodes $1$ or $m$, the components of the eigenstate of the largest eigenstate ($x_1$ or $x_m$) give large contribution to the amplitude, the time range of the scaling behavior is very short (See Fig.~\ref{fg3}(d)). In this case, the return probabilities do not decay ad infinitum but fluctuate about a constant value, suggesting significant localization induced by the largest eigenvalue. We will study this feature in detail in the following.

\subsection{Probability Distribution}
In this section, we consider the long time averaged probabilities. We find that these limiting probabilities display a symmetric structure, symmetric nodes $k$ and $l$ ($k+l=m+1$ or $k+l=N+m+1$, see the two nodes in the horizontal projection in Fig.~\ref{fg1}) have the same limiting probabilities. This suggests that the probability of finding the exciton at a certain node $k$ is equal to the probability of finding the exciton at the symmetric position $l=(m+1)-k$ (or $l=N+m+1-k$), {\em i.e.}, $\chi_{k,j}=\chi_{m+1-k,j}\ \forall \ j,k$ (or $\chi_{k,j}=\chi_{{\scriptscriptstyle N}+m+1-k,j}\ \forall \ j,k$). This symmetric structure can be understood as a result of the axis symmetry of graph. Here, we give a rigorous mathematical proof for this symmetric structure using Eq.~(\ref{eq4}). In Eq.~(\ref{eq4}), the limiting probabilities are only dependent on the eigenstates. Suppose the $n$th eigenstate is expanded as $|\Psi_n\rangle=\sum_{j=1}^Nx_j^{(n)}|j\rangle$, the probabilities in Eq.~(\ref{eq4}) can be written as $\chi_{k,j}=\sum_n|x_j^{(n)}|^2|x_k^{(n)}|^2$. In Appendix C, we prove that the components of eigenstate are symmetric, $|x_j|=|x_{m+1-j}|,\  \forall \ j\in [1,m]$; $|x_j|=|x_{m+{\scriptscriptstyle N}+1-j}|,\ \forall \ j\in [m+1,N]$. Here, no superscript for $x_j$ means the identity holds for all the eigenstates ($\forall \ n$). The symmetric structure of the eigenstates leads to the symmetric limiting probabilities. This is one of the main conclusions in our paper.
\begin{figure}
\scalebox{0.5}[0.5]{\includegraphics{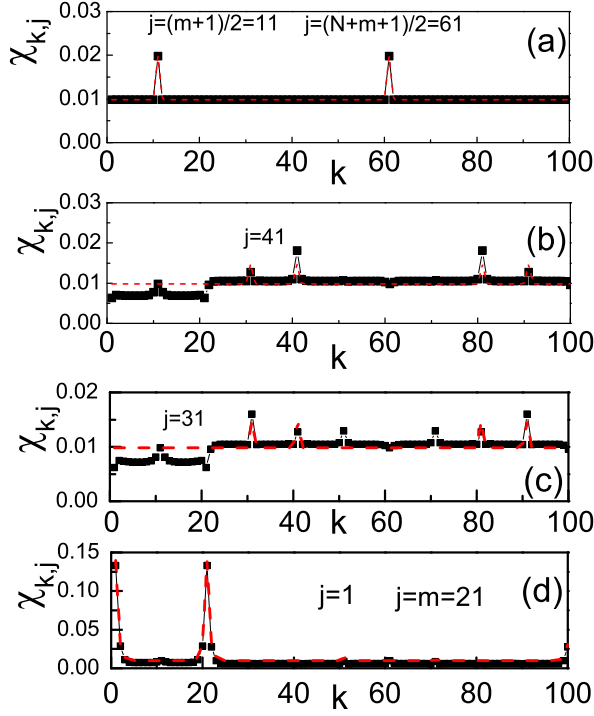}} \caption{(Color online) Long-time averaged probability distribution $\chi_{k,j}$ on a network of $N=100$ and $m=21$ for different starting positions $j$ (compare Fig.~\ref{fg3}). (a)The exciton starts at the center node between $1$ and $m$, {\em i.e.}, $j = (m +
1)/2 = 11$ and $j = (N +m+ 1)/2 = 61$. (b)Starting at j = 41. (c)Starting at j = 31. (d)Starting at j = 1 and j = m = 21. The black squares are the exact numerical results obtained by diagonalizing the Laplacian matrix. The dashed red lines are the analytical predictions in Eq.~(\ref{eq34}).
\label{fg4}}
\end{figure}

Now we try to use the eigenstates to find analytical approximate results for the limiting probabilities. The contribution in Eq.~(\ref{eq4}) comes from three parts: the eigenstate of the largest eigenvalue, non-degenerated eigenstates ($|\Psi_{0}^{(0)}\rangle$ and $|\Psi_{N/2}^{(0)}\rangle$) and degenerated eigenstates. Therefore Eq.~(\ref{eq4}) transforms into
\begin{equation}\label{eq33}
\begin{aligned}
\chi_{k,j}\approx &|x'_j|^2|x'_k|^2+\frac{1}{N^2}(1+\lambda) \\
&+\sum_{n=1}^{\lceil N/2\rceil-\lambda}|\langle k| \Psi_n^{(0)}\rangle |^2|\langle j|\Psi_n^{(0)}\rangle |^2 \\
&+\sum_{n=1}^{\lceil N/2\rceil-\lambda}|\langle k| \Psi_{-n}^{(0)}\rangle |^2|\langle j|\Psi_{-n}^{(0)}\rangle |^2,
\end{aligned}
\end{equation}
where $\lambda$ takes value $1$ for even $N$ and $0$ for odd $N$. Substituting the degenerated eigenstates of Eq.~(\ref{eq32}) into the above equation and noting the trigonometric function formula $\sum_{i=1}^j\cos\frac{2\pi ix}{N}=\frac{1}{2}(\sin\frac{(2j+1)\pi x}{N}/\sin\frac{\pi x}{N}-1)$, Eq.~(\ref{eq33})
can be simplified as,
\begin{equation}\label{eq34}
\begin{aligned}
\chi_{k,j}\approx &|x'_j|^2|x'_k|^2 +\frac{1}{N}- \frac{1}{N^2}\\
&+\frac{1}{2N^2}\bigg\{ \frac{\sin2\pi(j-k)(1-\frac{\lambda}{N})}{\sin\frac{2\pi(j-k)}{N}}  \\
&+\frac{\sin2\pi(j+k-m-1)(1-\frac{\lambda}{N})}{\sin\frac{2\pi(j+k-m-1)}{N}} \bigg\}
\end{aligned}
\end{equation}

Fig.~\ref{fg4} shows the probability distribution for different starting positions. It is evident that the exact numerical
results agree well with analytical predictions in Eq.~(\ref{eq34}). For exciton starting at $j=1$ and $j=m$, there is a high probability to find the exciton at $j=1$ and $j=m$, suggesting significant localization in the dynamics. For excitons starting at center node $j=(m+1)/2$ ($m$ is odd) and $j=(N+m+1)/2$ ($N+m$ is odd), the probability distribution is exact the same as the cycles. This could be explained by Eq.~(\ref{eq34}) as follows: For $j=(m+1)/2$ or $j=(N+m+1)/2$, $|x'_j|=0$ (See Eqs. (\ref{c13})-(\ref{c14}) in Appendix C), the contribution from the largest eigenstate's eigenstate is $0$ (See the first term in Eq.~(\ref{eq34})), the localized probability $\chi_{k,j}$ mainly depends on the trigonometric function in Eq.~(\ref{eq34}). For return probabilities with $j=k=(m+1)/2$ ($m$ is odd) or $j=k=(N+m+1)/2$ ($N+m$ is odd), limits of the trigonometric function ratio in the big bracket equal to $N-\lambda$, thus Eq.~(\ref{eq34}) becomes as,
\begin{equation}\label{eq35}
\chi_{k,j}=\left\{
\begin{array}{ll}
\frac{(2N-1-\lambda)}{N^2},   &  \ j=k=\frac{m+1}{2},\frac{(N+m+1)}{2} \\
\frac{(N-1-\lambda)}{N^2},   &  \ j\neq k,\ \ j,k=\frac{m+1}{2}, \frac{(N+m+1)}{2} \\
\end{array}
\right.
\end{equation}
which are consistent with the results in Ref.~\cite{rn21}. For this particular case, $j,k=\{ \frac{m+1}{2}, \frac{(N+m+1)}{2}\}$, the probability distribution is exact the same as the cycles, this indicates the adding of link has no impact on the dynamics for center node excitations ($k$ or $j$ at the symmetry axis, See Fig.~\ref{fg1}). Here we provide a theoretical interpretation for this phenomena. However, if $m$ and $N+m$ are not odd numbers, this particular case does not exist.

\begin{figure}
\scalebox{0.33}[0.33]{\includegraphics{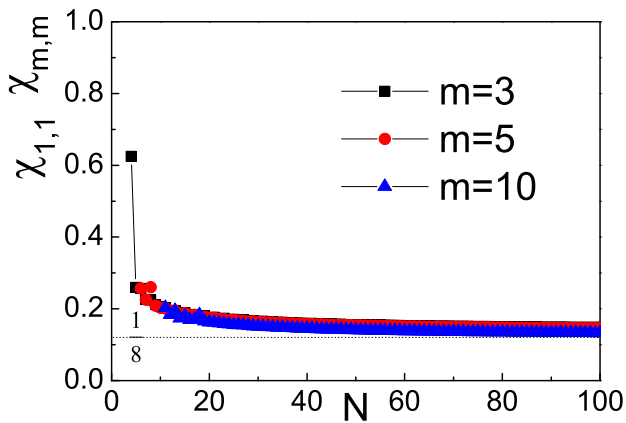}} \caption{(Color online) Return probabilities $\chi_{1,1}$ and $\chi_{m,m}$ as a function of the network size $N$ for $m=3$ (black squares), $m=5$ (red dots) and $m=10$ (blue triangles). All the return probabilities approach to the predicted constant value $1/8$ (The horizontal dotted line).
\label{fg5}}
\end{figure}

As we have shown, the return probabilities $\chi_{j,j}$ has a high value at $j=1$ and $j=m$. Such a strong localization may vanish as the size of the system becomes large. In order to investigate the dependence of localized probability $\chi_{1,1}$ (or $\chi_{m,m}$) on the size of the system, we plot $\chi_{1,1}$ (or $\chi_{m,m}$) as a function of $N$ for different values of $m$ in Fig.~\ref{fg5}. We find that the return probability converges to a constant value when $N$ tends to infinity. This constant value can be calculated using Eq.~(\ref{eq34}). For infinite systems, $\chi_{1,1}$ (or $\chi_{m,m}$) equals to the first term of Eq.~(\ref{eq34}) (the other terms are $0$ in the limit of $N\to \infty$). Therefore, for infinite systems, the limiting probabilities are determined by the components of the eigenstate of the largest eigenvalue. Noting that $|x'_1|=|x'_m|=2^{-3/4}$, we get
\begin{equation}\label{eq36}
\chi_{1,1}=\chi_{1,m}=\chi_{m,1}=\chi_{m,m}\approx |x'_1|^4=\frac{1}{8}
\end{equation}
Therefore $\chi_{1,1}$ (or $\chi_{m,m}$) converges to the $1/8$, indicating there is a significant lower bound for the return probabilities. This result has not been previously reported and suggests that quantum walks display nontrivial localization on nonsymmetric system. Since all the components of the largest eigenvalue's eigenstate are known (See Eqs.~(\ref{eq25}) and (\ref{eq26})), the limiting probabilities $\chi_{k,j}$ have a lower bound given by $|x'_j|^2|x'_k|^2$. Using the expression of $x'_j$ in Eq.~(\ref{eq26}), $|x'_j|^2|x'_k|^2\approx |z_0|^{-2(d_j+d_k)}|x'_m|^4=\frac{1}{8}|z_0|^{-2(d_j+d_k)}$. The lower bound for the limiting probability is determined by the shortest distance $d_j$ and $d_k$.
For instance, for $j=1$ and $k=2$, $d_j=0$, $d_k=1$, $\chi_{1,2}>\frac{1}{8(1+\sqrt{2})^2}$; for $j=m$ and $k=N-1$, $d_j=0$, $d_k=2$, $\chi_{k,j}>\frac{1}{8(1+\sqrt{2})^4}$, {\em etc.} This dynamical feature is completely different from the case of the cycles. For continuous-time quantum walks on a 1D cycles of $N$ nodes, all the limiting probabilities $\chi_{k,j}$ are $0$ when $N\to \infty$, due to all the components of the eigenstates
approaches to $0$ in this case. So there is no localization on infinite regular cycles. For our case, the localization is induced by the largest eigenvalue and size of the system $N$, which differs from the 1D regular cycles where the localization is only induced by the size of the system. To this end, the localization in our case is much more essential and may have potential applications in quantum information science.
\section{Trapping}
An important process related to random walk is
trapping~\cite{rn22,rn23}. Trapping problems have been widely
studied in the frame of physical chemistry, as part of the general
reaction-diffusion scheme~\cite{rn24}. Previous work has been
devoted to the trapping problem on discrete-time random
walks~\cite{rn25,rn26}. However, even in its simplest form, trapping
was shown to yield a rich diversity of results, with varying
behavior over different geometries, dimension, and time
regimes~\cite{rn26}. The main physical quantity related to trapping
process is the survival probability, which denotes the probability
that a particle survives during the walk in a space with traps.

In this paper, we consider trapping using the approach based on time
dependent perturbation theory and adopt the methodology proposed in
Ref.~\cite{rn27}. In Ref.~\cite{rn27}, the authors consider a system
of $N$ nodes and among them $M$ are traps ($M<N$). The trapped nodes
are denoted them by $m'$, so that $m'\in {\cal M}$. The new
Hamiltonian of the system is $H=H_0+i\bm{\Gamma}$, where $H_0$ is the
original Hamiltonian without traps and $i\Gamma$ is the trapping
operator. $\bm{ \Gamma}$ has $m$ purely imaginary diagonal elements
$\bm{\Gamma}_{m'm'}$ at the trap nodes and assumed to be equal for all $m$
($\bm{\Gamma}_{m'm'}\equiv \Gamma >0$). See Ref.~\cite{rn27} for details.
The new Hamiltonian is non-hermitian and has $N$ complex eigenvalues
and eigenstates \{$E_l$, $|\Psi_l\rangle$\} ($l=1,2,...,N$). Then
the quantum transition probability is
\begin{equation}\label{eq37}
\pi_{k,j}(t)=|\alpha_{k,j}(t)|^2=|\sum_l e^{-itE_l}\langle k
|\Psi_l\rangle \langle \tilde{\Psi}_l|j\rangle|^2,
\end{equation}
where $\langle \tilde{\Psi}_l|$ ($l=1,2,...,N$) is the conjugate
eigenstates of the new Hamiltonian. Equation~(\ref{eq37}) depends on the initially excited node $j$. The
average survival probability over all initial nodes $j$ and all
final nodes $k$, neither of them being a trap node, is given by~\cite{rn27},
\begin{equation}\label{eq38}
\Pi_{M}(t)=\frac{1}{N-M}\sum_{j\not{\in} {\cal M}}\sum_{k\not{\in}
{\cal M}}\pi_{k,j}(t).
\end{equation}
For intermediate and long times and a small number of trap nodes,
$\Pi_{M}(t)$ is mainly a sum of exponentially decaying terms~\cite{rn27}:
\begin{equation}\label{eq39}
\Pi_{M}(t)\approx\frac{1}{N-M}\sum_{l=1}^N\exp({-2\gamma_lt}),
\end{equation}
where $\gamma_l$ is the imaginary part of the eigenvalue $E_l$.

\begin{figure}
\scalebox{0.37}[0.37]{\includegraphics{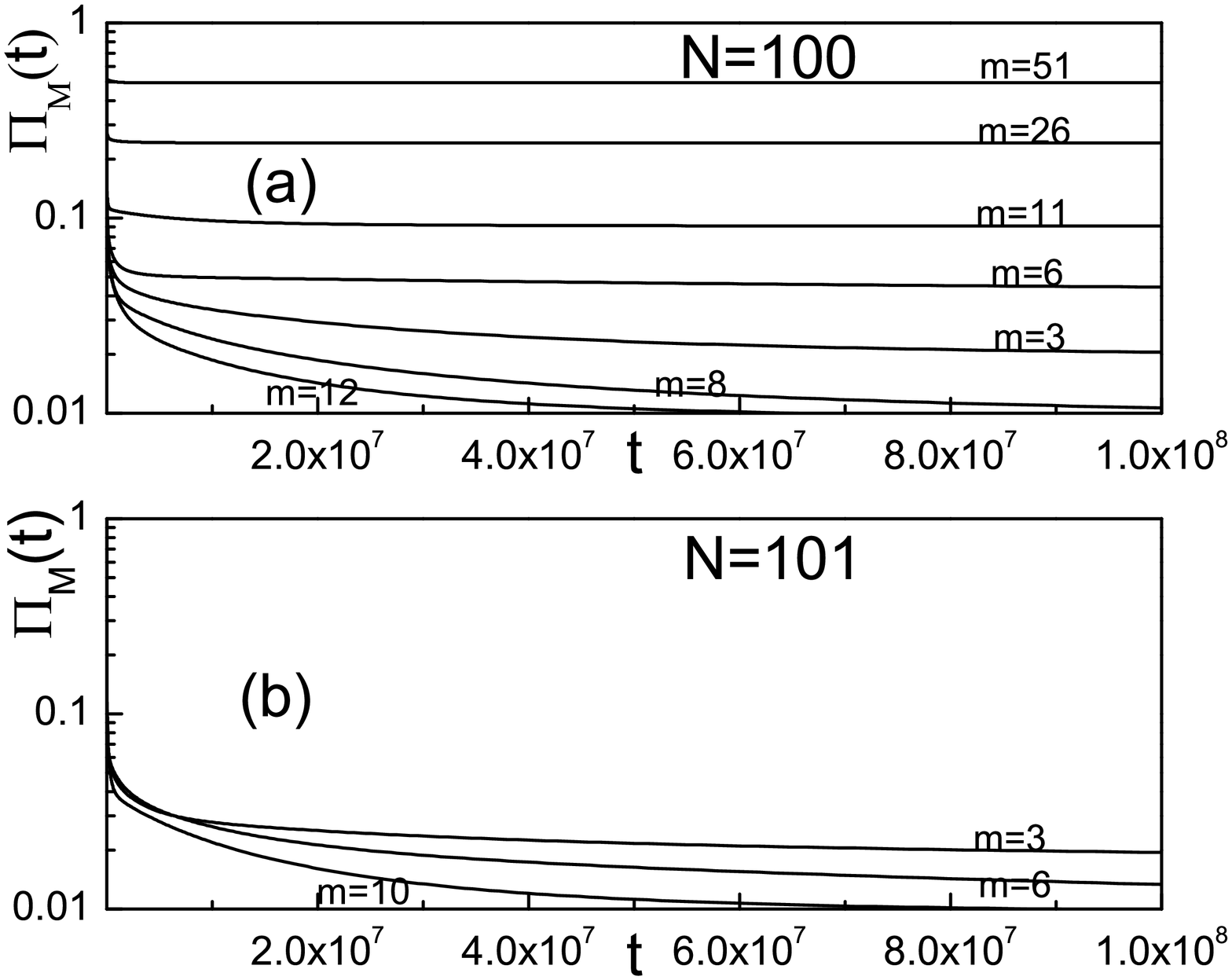}} \caption{Survival probabilities $\Pi_M(t)$ for $N=100$ (a) and $N=101$ (b) for different values of $m$. In the calculation, the trap is localized at node $1$ and $\Gamma=1$. We find that if $\frac{N}{2(m-1)}$ is integral, $\Pi_M(t)$ converges to $(m-1-\lambda)/(N-1)$. Otherwise, $\Pi_M(t)$ approaches to $0$.
\label{fg6}}
\end{figure}

For the cycle with an additional link, we focus on the case that only one trap exists in the graph and the trap is located on node $1$.
Fig.~\ref{fg6} shows the survival probability $\Pi_M(t)$ vs $t$ for $N=100$ (a) and $N=101$ (b). We find that $\Pi_M(t)$ decays very slowly and converges to
a certain constant value (nonzero) for some special network parameters $N$ and $m$. Meanwhile, for the other values of $N$ and $m$, $\Pi_M(t)$ decays slowly and converges to $0$. For the network of size $N=100$ (See Fig.~\ref{fg6}(a)), $\Pi_M(t)$ converges to nonzero constants for $m=3$, $m=6$, $m=11$, $m=26$ and $m=51$, whereas $\Pi_M(t)$ approaches to $0$ for other values of $m$. For the network of size $N=101$ (See Fig.~\ref{fg6}(b)), $\Pi_M(t)$ approaches to $0$ for all values of $m$. The nonzero threshold of the survival probability suggests that the trapping process displays significant localization
for some special values of network parameters. This result is interesting since in previous studies the survival probability converges to $0$~\cite{rn27,rn11}, indicating the excitation eventually returns to the traps. The nonzero threshold of the survival probability implies that the excitation does not eventually returns to the traps~\cite{add7}, and the survival probability is smaller than $1/2$.

To give an explanation to the localized survival probability, we analysis the Laplacian eigenvalues using the perturbation theory. In Eq.~(\ref{eq39}), the behavior of $\Pi_M(t)$ is determined by the the imaginary part of the eigenvalue ($\gamma_l$). If $\gamma_l$ is nonzero ($\gamma_l>0$), $\exp(-2\gamma_lt)$ converges to $0$ at long time scale. On the contrary, if $\gamma_l$ is $0$, $\exp(-2\gamma_lt)$ equals to $1$. The main contribution for the localized survival probability comes from the real eigenvalues ($\gamma_l=0$), thus the lower bound of $\Pi_M(t)$ depends on the number of real eigenvalues ($\gamma_l=0$). The eigenvalues and eigenstates for cycle with an additional link has been analyzed in Sec. III, the trapping operator can be regarded as the perturbed operator. According to perturbation theory, the first order corrections of the eigenvalues of the non-hermitian Hamiltonian are given by
\begin{equation}\label{eq40}
E_n^{(1)}= \langle \Psi_n^{(0)}|i\bm{\Gamma}|\Psi_n^{(0)}\rangle=i\Gamma|\langle 1|\Psi_n^{(0)}\rangle|^2.
\end{equation}
Utilizing the zero order approximation of the eigenstates in Eq.~(\ref{eq32}), the first order correction of the eigenvalue can be written as $E_n^{(1)}=i\Gamma\frac{1}{N}[1\pm\cos(m-1)\theta_n]$. $\gamma_l=0$ requires the first order correction $E_n^{(1)}$ equals to $0$, thus $(m-1)\theta_n=\frac{2n(m-1)\pi}{N}=k\pi$ ($n\in[1,\lceil N/2\rceil-\lambda]$, $k>0$). The number of integer solutions of this equation gives number of real eigenvalues $\gamma_l=0$. If $\frac{N}{2(m-1)}$ is integral, the number of integer solutions is $(m-1-\lambda)$. Therefore, the lower bound of $\Pi_M(t)$ equals to $(m-1-\lambda)/(N-1)$ ($<\frac{1}{2}$). If $N$ is odd, no integer solutions exist, thus $\Pi_M(t)$ converges to $0$ for odd size of networks. Thus we succeed in explaining the localized survival probability.
\section{CONCLUSIONS AND DISCUSSIONS}
In summary, we consider the continuous-time quantum walk on the cycle with an additional link. We analytically
treat this problem and  and approximate the Laplacian eigenvalues and eigenstates by the Chebyshev polynomial technique and perturbation theory for the first time. We find that the probability evolution exhibits a similar behavior like the cycle if the exciton starts far away from the two ends of the added link. The distribution of the long-time limiting probabilities display symmetric structure, we prove this symmetry using the
exact determinant equation for the eigenvalues expressed by Chebyshev polynomials. In addition, the
quantum dynamics exhibit significant localization when the walk starts at the two ends of the extra
link, we show that the localized probability is determined by the largest Laplacian eigenvalue and
there is a significant lower bound for it even in the limit of infinite system. Finally, we study the problem of trapping and show the survival probability also displays significant localization for some special values of network parameters, we succeed
in explaining this phenomenon and determining the conditions for the emergence of such localization using the perturbation theory.

In our work, we find that the extra link of the cycle indeed cause a different dynamical behavior compared to the dynamics on the cycle. The impact of the extra link is mainly determined by the largest eigenvalue and its corresponding eigenstate. This is similar to the other dynamic processes on networks where the largest eigenvalue plays an important role in relevant dynamics. The additional link causes a significant localization at the two end nodes of the link. We show such localization is related to the eigenstate of the largest eigenvalue, which differs from the 1D regular cycles where the
localization is only induced by the size of the system. This characteristic is reminiscent of the Anderson localization in condensed matter physics~\cite{add5}. Anderson has shown that there is no quantum diffusion on disordered medium, an effect nowadays called (strong) localization, where the
quantum mechanical transport through the lattice is prohibited by disordered structures~\cite{add5,add6}. In our case, the additional link breaks the perfect order nature of the cycle and induces some structural disorder, this also leads to a strong localization resembling the Anderson localization. For Anderson localization all the eigenstates become localized, whereas here it is only the one corresponding to the largest eigenvalue. We hope our findings provide a deeper understanding for the dynamics of quantum walks on networks and possible insights into the experimental implementation of continuous-time quantum walks on irregular networks~\cite{rn28}.
\begin{acknowledgments}
This work is partially supported by Heiwa Nakajima Foundation of Japan, National Natural Science
Foundation of China under projects 11079027, Doctor Fund Project of Ministry of Education under Contract 20103201120003 and Shanghai Key Laboratory
of Intelligent Information Processing (IIPL-2011-009). Yusuke Ide is supported by the Grant-in-Aid for Young Scientists
(B) of Japan Society for the Promotion of Science (Grant No. 23740093). Norio Konno is supported by the Grant-in-Aid for Scientific Research (C) of Japan
Society for the Promotion of Science (Grant No. 21540118).
\end{acknowledgments}
\appendix
\onecolumngrid
\section{Definition of Chebyshev polynomials}
The Chebyshev polynomials of the first kind are defined by the recurrence relation~\cite{rn29,rn30},
\begin{equation}\label{a1}
T_0(x)=1, T_1(x)=x, 2xT_n(x)=T_{n-1}(x)+T_{n+1}(x).
\end{equation}
The Chebyshev polynomials of the second kind are defined by the recurrence relation~\cite{rn29,rn30},
\begin{equation}\label{a2}
U_0(x)=1, U_1(x)=2x, 2xU_n(x)=U_{n-1}(x)+U_{n+1}(x).
\end{equation}
The closed-form solutions of Eqs.~(\ref{a1}) and (\ref{a2}) are given by,
\begin{align}
T_n(x)&=\frac{z^n+z^{-n}}{2}, \label{a3} \\
U_n(x)&=\frac{z^{-(n+1)}-z^{n+1}}{|z^{-1}-z|}, \label{a4}
\end{align}
where $z=x-\sqrt{x^2-1}$. For the case of large order $n$ and $z<-1$ ($|z|>1$), the above solutions can be approximated by,
\begin{equation}\label{a5}
T_n(x)\approx\frac{z^n}{2}, U_n(x)\approx-\frac{z^{n+1}}{|z^{-1}-z|}=-\frac{z^{n+1}}{z^{-1}-z}
\end{equation}
Using the closed-form solutions for the Chebyshev polynomials (See Eqs.~(\ref{a3}) and (\ref{a4})), we can prove the following identities,
\begin{eqnarray}
U_{n-1}(x)+U_{-n-1}(x)=0,  \label{a6} \\
2xU_n(x)=U_{n-1}(x)+U_{n+1}(x), \label{a7}\\
T_n(x)=U_n(x)-xU_{n-1}(x)=xU_{n-1}(x)-U_{n-2}(x), \label{a8} \\
U_n(x)U_m(x)-U_{n-1}(x)U_{m-1}(x)=U_{n+m}(x), \label{a9}   \\
U_n(x)U_m(x)-U_{n+1}(x)U_{m-1}(x)=U_{n-m}(x), \label{a10} \\
U_n(x)T_m(x)+U_{m-1}(x)T_{n+1}(x)=U_{n+m}(x), \label{a11}  \\
U_n(x)T_m(x)-U_{m-1}(x)T_{n+1}(x)=U_{n-m}(x), \label{a12} \\
T_n^2(x)-(x^2-1)U_{n-1}^2(x)=1,  \label{a13} \\
T_m(x)U_n(x)=\frac{1}{2}[U_{m+n}(x)+U_{n-m}(x)], \ \ T_m(x)T_n(x)=\frac{1}{2}[T_{m+n}(x)+T_{|m-n|}(x)]  . \label{a14}
\end{eqnarray}

\section{Calculation for the determinant equation}
For the sake of simplicity, we first simplify the four coefficients $c_1$, $c_2$, $c_3$ and $c_4$ using the Chebyshev polynomial identities. The first coefficient $c_1$ can be simplified as,
\begin{equation}\label{b1}
\begin{aligned}
c_1&=U_{{\scriptscriptstyle N}-m}(x)[(2x+1)U_{m-2}(x)-U_{m-3}(x)]
-U_{{\scriptscriptstyle N}-m-1}(x)U_{m-2}(x) \\
&=\{  U_{{\scriptscriptstyle N}-m}(x)U_{m-1}(x)-U_{{\scriptscriptstyle N}-m-1}(x)U_{m-2}(x) \}
+U_{{\scriptscriptstyle N}-m}(x)U_{m-2}(x) \ \ \ \ \ (\ref{a7}) \ \text{used}\\
&=U_{{\scriptscriptstyle N}-1}(x)+U_{{\scriptscriptstyle N}-m}(x)U_{m-2}(x)\ \ \ \ \ \ \ \ \ \ \ \ \ \ \ \ \ \ \ \ \ \ \ \ \ \ \ \ \ \ \ \ \ \ \ \ \ \ \ \ \ \ \ \ \ \ \ \ \ \ \ (\ref{a9}) \ \text{used}
\end{aligned}
\end{equation}
The second coefficient can be simplified as,
\begin{equation}\label{b2}
\begin{aligned}
c_2&=U_{{\scriptscriptstyle N}-m}(x)[U_{m-4}(x)-(2x+1)U_{m-3}(x)-1]
+U_{{\scriptscriptstyle N}-m-1}(x)U_{m-3}(x)-1  \\
&=[  -U_{{\scriptscriptstyle N}-m}(x)U_{m-2}(x)+U_{{\scriptscriptstyle N}-m-1}(x)U_{m-3}(x)  ]
-U_{{\scriptscriptstyle N}-m}(x)U_{m-3}(x)-U_{{\scriptscriptstyle N}-m}(x)-1 \ \ \ (\ref{a7}) \ \text{used}\\
&=-U_{{\scriptscriptstyle N}-2}(x)-U_{{\scriptscriptstyle N}-m}(x)U_{m-3}(x)-U_{{\scriptscriptstyle N}-m}(x)-1,  \ \ \ \ \ \ \ \ \ \ \ \ \ \ \ \ \ \ \ \ \ \ \ \ \ \ \ \ \ \ \ \ \ \ \ \ \ \  \ \ \ \ \ \ \ \ (\ref{a9}) \ \text{used}
\end{aligned}
\end{equation}
Analogously, $c_3$ and $c_4$ can be simplified as,
\begin{equation}\label{b3}
\begin{aligned}
c_3&=U_{{\scriptscriptstyle N}-m-1}(x)[(2x+1)U_{m-2}(x)-U_{m-3}(x)]
-U_{{\scriptscriptstyle N}-m-2}(x)U_{m-2}(x)+U_{m-2}(x)+1 \\
&=\{  U_{{\scriptscriptstyle N}-m-1}(x)U_{m-1}(x)-U_{{\scriptscriptstyle N}-m-2}(x)U_{m-2}(x)  \}
 +U_{{\scriptscriptstyle N}-m-1}(x)U_{m-2}(x)+U_{m-2}(x)+1 \ \ \ \ \\
&=U_{{\scriptscriptstyle N}-2}(x)+U_{{\scriptscriptstyle N}-m-1}(x)U_{m-2}(x)+U_{m-2}(x)+1,
\end{aligned}
\end{equation}
\begin{equation}\label{b4}
\begin{aligned}
c_4&=U_{{\scriptscriptstyle N}-m-1}(x)[U_{m-4}(x)-(2x+1)U_{m-3}(x)-1]
+U_{{\scriptscriptstyle N}-m-2}(x)U_{m-3}(x)-U_{m-3}(x)-(2x+1) \\
&=\{ -U_{{\scriptscriptstyle N}-m-1}(x)U_{m-2}(x)+U_{{\scriptscriptstyle N}-m-2}(x)U_{m-3}(x)  \}
-U_{{\scriptscriptstyle N}-m-1}(x)U_{m-3}(x)-U_{{\scriptscriptstyle N}-m-1}(x)-U_{m-3}(x)-(2x+1) \\
&=-U_{{\scriptscriptstyle N}-3}(x)-U_{{\scriptscriptstyle N}-m-1}(x)U_{m-3}(x)
-U_{{\scriptscriptstyle N}-m-1}(x)-U_{m-3}(x)-(2x+1)
\end{aligned}
\end{equation}
Substitute the simplified $c_1$, $c_2$, $c_3$ and $c_4$ into $c_1c_4-c_2c_3$ and expand the two products, we obtain,
\begin{equation}\label{b5}
\begin{split}
c_1c_4-c_2c_3=&1+U_{{\scriptscriptstyle N}-m}(x)+ U_{m-2}(x)-U_{{\scriptscriptstyle N}-1}(x)
+\{ U_{{\scriptscriptstyle N}-2}^2(x)-U_{{\scriptscriptstyle N}-1}(x)U_{{\scriptscriptstyle N}-3}(x) \}  \\
&+\{ U_{{\scriptscriptstyle N}-2}(x)U_{{\scriptscriptstyle N}-m}(x)-U_{{\scriptscriptstyle N}-1}(x)U_{{\scriptscriptstyle N}-m-1}(x) \}
+\{ U_{{\scriptscriptstyle N}-2}(x)U_{m-2}(x)-U_{{\scriptscriptstyle N}-1}(x)U_{m-3}(x) \}  \\
&+\{ U_{{\scriptscriptstyle N}-2}(x)U_{m-3}(x)-U_{{\scriptscriptstyle N}-3}(x)U_{m-2}(x) \}U_{{\scriptscriptstyle N}-m}(x)
+\{ U_{{\scriptscriptstyle N}-2}(x)U_{m-2}(x)-U_{{\scriptscriptstyle N}-1}(x)U_{m-3}(x) \}U_{{\scriptscriptstyle N}-m-1}(x) \\
&+ [ U_{{\scriptscriptstyle N}-m}(x)U_{m-3}(x)+U_{{\scriptscriptstyle N}-m-1}(x)U_{m-2}(x)
-2xU_{{\scriptscriptstyle N}-m}(x)U_{m-2}(x)]+2[ U_{{\scriptscriptstyle N}-2}(x)-xU_{{\scriptscriptstyle N}-1}(x)].
\end{split}
\end{equation}
In the above equation, the terms in the big brackets $\{\ \ \}$ can be simplified using (\ref{a10}), and the terms in the middle brackets $[\ \ ]$ can be simplified using (\ref{a7}) and (\ref{a8}). Thus,
\begin{equation}\label{b6}
\begin{aligned}
c_1c_4-c_2c_3=&2+U_{{\scriptscriptstyle N}-m}(x)+ 2U_{m-2}(x)-U_{{\scriptscriptstyle N}-1}(x)
+ U_{{\scriptscriptstyle N}-m}(x)
-U_{{\scriptscriptstyle N}-m-1}(x)U_{{\scriptscriptstyle N}-m}(x)
+U_{{\scriptscriptstyle N}-m}(x)U_{{\scriptscriptstyle N}-m-1}(x) \\
&+ [ U_{{\scriptscriptstyle N}-m}(x)U_{m-3}(x)+U_{{\scriptscriptstyle N}-m-1}(x)U_{m-2}(x)
-(U_{{\scriptscriptstyle N}-m-1}(x)+U_{{\scriptscriptstyle N}-m+1}(x))U_{m-2}(x)]-2T_{{\scriptscriptstyle N}}(x).
\end{aligned}
\end{equation}
Applying identity (\ref{a9}) to the term in the middle bracket leads to,
\begin{equation}\label{b7}
c_1c_4-c_2c_3=2\bigg\{ 1+U_{{\scriptscriptstyle N}-m}(x)+ U_{m-2}(x)-U_{{\scriptscriptstyle N}-1}(x)-T_{{\scriptscriptstyle N}}(x) \bigg\}
\end{equation}

Using the Chebyshev polynomial identity (\ref{a7}) in Appendix A, Eq.~(\ref{eq16}) can be written as $x_{\scriptscriptstyle N}=[U_{m-1}(x)+U_{m-2}(x)]x_{m-1}-[U_{m-2}(x)+U_{m-3}(x)+1]x_m$. According to Eq.~(\ref{eq18}), $x_{m-1}=-c_2x_m/c_1$ and noting the simplified $c_1$, $c_2$ in (\ref{b1}) and (\ref{b2}), thus we can write $x_{\scriptscriptstyle N}$ as,
\begin{equation}
\begin{aligned}\label{b8}
x_{\scriptscriptstyle N}&=-\big( U_{m-1}(x)+U_{m-2}(x) \big) \frac{c_2}{c_1}x_m-\big( U_{m-2}(x)+U_{m-3}(x)+1  \big) x_m\\
&=\bigg\{   \frac{U_{{\scriptscriptstyle N}-2}(x)+U_{{\scriptscriptstyle N}-m}(x)U_{m-3}(x)+U_{{\scriptscriptstyle N}-m}(x)+1}{U_{{\scriptscriptstyle N}-1}(x)+U_{{\scriptscriptstyle N}-m}(x)U_{m-2}(x)}\big( U_{m-1}(x)+U_{m-2}(x)\big) -\big( U_{m-2}(x)+U_{m-3}(x)+1 \big)  \bigg\}x_m \\
&=\bigg\{   \frac{  U_{{\scriptscriptstyle N}-m}(x)[U_{m-1}(x)U_{m-3}(x)-U^2_{m-2}(x)]+[U_{{\scriptscriptstyle N}-2}(x)U_{m-1}(x)-U_{{\scriptscriptstyle N}-1}(x)U_{m-2}(x)]   }{U_{{\scriptscriptstyle N}-1}(x)+U_{{\scriptscriptstyle N}-m}(x)U_{m-2}(x)} \\
&\ \ \ +\frac{[U_{{\scriptscriptstyle N}-2}(x)U_{m-2}(x)-U_{{\scriptscriptstyle N}-1}(x)U_{m-3}(x)]+U_{{\scriptscriptstyle N}-m}(x)U_{m-1}(x)+U_{m-1}(x)+U_{m-2}(x)-U_{{\scriptscriptstyle N}-1}(x)  }
{U_{{\scriptscriptstyle N}-1}(x)+U_{{\scriptscriptstyle N}-m}(x)U_{m-2}(x)} \bigg\}x_m \\
&=\frac{-U_{{\scriptscriptstyle N}-m}(x)+U_{{\scriptscriptstyle N}-m-1}(x)+U_{{\scriptscriptstyle N}-m}(x)+  U_{{\scriptscriptstyle N}-m}(x)U_{m-1}(x)+U_{m-1}(x)+U_{m-2}(x)-U_{{\scriptscriptstyle N}-1}(x) }{U_{{\scriptscriptstyle N}-1}(x)+U_{{\scriptscriptstyle N}-m}(x)U_{m-2}(x)}x_m\\
&=\frac{U_{{\scriptscriptstyle N}-m-1}(x)+  U_{{\scriptscriptstyle N}-m}(x)U_{m-1}(x)+U_{m-1}(x)+U_{m-2}(x)-U_{{\scriptscriptstyle N}-1}(x) }{U_{{\scriptscriptstyle N}-1}(x)+U_{{\scriptscriptstyle N}-m}(x)U_{m-2}(x)}x_m\equiv\frac{c'_2}{c_1} x_m,
\end{aligned}
\end{equation}
where the terms in the middle brackets ($[\ \ ]$) are simplified using (\ref{a10}).

\section{Proof for the symmetry of the eigenstates}
In this section, we prove the symmetry of the components of the eigenstates, {\em i.e.},
\begin{equation}\label{c1}
|x_j|=\begin{cases}
\ |x_{m+1-j}|, & \text{if } j\in [1,m] \\
\ |x_{m+{\scriptscriptstyle N}+1-j}|, & \text{if } j\in [m+1,N]
\end{cases}
\end{equation}
To prove the symmetric behavior of the eigenstates, we have the following theorem:
If $x_1=\pm x_m$ is true, then all the other symmetric relations are true, {\em i.e.},
\begin{equation}\label{c2}
\begin{aligned}
\text{If } &x_1=\pm x_m \ \text{is true, then} \
 x_2=\pm x_{m-1}, x_{\scriptscriptstyle N}=\pm x_{m+1},...\  \text{are also true}.
\end{aligned}
\end{equation}
In the above theorem, the signs $\pm$ are uniformed. This theorem can be proved using Eq.~(\ref{eq12}) and the Chebyshev identity. Set $j=2$ and $j=3$ in Eq.~(\ref{eq12}), we obtain,
\begin{align}
x_1=U_{m-2}(x)x_{m-1}-U_{m-3}(x)x_m, \label{c3} \\
x_2=U_{m-3}(x)x_{m-1}-U_{m-4}(x)x_m. \label{c4}
\end{align}
If $x_1=\pm x_m$, then $x_m=U_{m-2}(x)x_{m-1}/(U_{m-3}(x)\pm 1)$ (Eq.~(\ref{c3})). Substituting $x_m$ into Eq.~(\ref{c4}), we get the relationship between $x_2$ and $x_{m-1}$,
\begin{equation}\label{c5}
\begin{aligned}
\frac{x_2}{x_{m-1}}&=U_{m-3}(x)-\frac{U_{m-2}(x)U_{m-4}(x)}{U_{m-3}(x)\pm  1}
=\frac{[U^2_{m-3}(x)-U_{m-2}(x)U_{m-4}(x)]\pm U_{m-3}(x)}{U_{m-3}(x)\pm  1}
=\frac{1\pm U_{m-3}(x)}{U_{m-3}(x)\pm  1}=\pm 1,
\end{aligned}
\end{equation}
where the term in the middle brackets ($[\ \ ]$) is simplified using (\ref{a10}). Thus we have proved $x_2=\pm x_{m-1}$. The remaining symmetric relations can also be proved in the same way (just use Eqs.~(\ref{eq12}) and (\ref{eq13})). The theorem suggests that to prove the symmetry of the eigenstates in Eq.~(\ref{c1}), we only need to prove one symmetric component of the eigenstates ($x_1$ and $x_m$) satisfies the Eq.~(\ref{c1}). In the following, we will try to prove $x_1=\pm x_m$ using the determinant equation (\ref{eq22}).

Now, we show that the determinant equation (\ref{eq22}) can be written as a product of two factors using the Chebyshev polynomial technique. According to (\ref{a12}), the term $U_{{\scriptscriptstyle N}-m}(x)$ can be written as $U_{{\scriptscriptstyle N}-m}(x)=U_{{\scriptscriptstyle N}-1}(x)T_{m-1}(x)-U_{m-2}(x)T_{\scriptscriptstyle N}(x)$. Substituting $U_{{\scriptscriptstyle N}-m}(x)$ into the determinant equation (\ref{eq22}), we get
\begin{equation}\label{c6}
U_{{\scriptscriptstyle N}-1}(x)T_{m-1}(x)=U_{m-2}(x)[T_{\scriptscriptstyle N}(x)-1]+T_{\scriptscriptstyle N}(x)+U_{{\scriptscriptstyle N}-1}(x)-1
\end{equation}
Square both sides of the above equation and note that $T^2_{m-1}(x)=(x^2-1)U^2_{m-2}(x)+1$ (Pell Equation in (\ref{a13})), we obtain,
\begin{equation}\label{c7}
\begin{split}
[(x^2-1)U^2_{{\scriptscriptstyle N}-1}(x)-(T_{\scriptscriptstyle N}(x)-1)^2]U^2_{m-2}(x)
=&2[T_{\scriptscriptstyle N}(x)-1][T_{\scriptscriptstyle N}(x)+U_{{\scriptscriptstyle N}-1}(x)-1]U_{m-2}(x) \\
&+[T_{\scriptscriptstyle N}(x)-1][T_{\scriptscriptstyle N}(x)+2U_{{\scriptscriptstyle N}-1}(x)-1]
\end{split}
\end{equation}
The factor in the middle bracket $[\ \ ]$ in the left hand side can be further simplified as,
\begin{equation}\label{c8}
\begin{split}
[(x^2-1)U^2_{{\scriptscriptstyle N}-1}(x)-(T_{\scriptscriptstyle N}(x)-1)^2]&=T^2_{\scriptscriptstyle N}(x)-1-(T_{\scriptscriptstyle N}(x)-1)^2
=2[T_{\scriptscriptstyle N}(x)-1]. \ \ \ \ \ \ \ \ \ \ \ \ (\ref{a13})\ \ \text{used}
\end{split}
\end{equation}
Therefore, Eq.~(\ref{c7}) becomes as,
\begin{equation}\label{c9}
[T_{\scriptscriptstyle N}(x)-1] \bigg\{   2U^2_{m-2}(x)-2[T_{\scriptscriptstyle N}(x)+U_{{\scriptscriptstyle N}-1}(x)-1]U_{m-2}(x)-[T_{\scriptscriptstyle N}(x)+2U_{{\scriptscriptstyle N}-1}(x)-1]     \bigg\}\equiv [T_{\scriptscriptstyle N}(x)-1]\Theta(N,m,x) =0
\end{equation}
In the above equation, $[T_{\scriptscriptstyle N}(x)-1]=0$ corresponds to the determinant equation for the cycles. Solving $[T_{\scriptscriptstyle N}(x)-1]=0$ gives the eigenvalues for the regular cycle of $N$ nodes, this suggests that some eigenvalues of the cycle are exactly the same as the eigenvalues in our model. Because we have squared the Eq.~(\ref{c6}), this doubles the number of solutions of the original determinant equation (\ref{eq22}). $[T_{\scriptscriptstyle N}(x)-1]=0$ only gives about one half ($\lceil N/2\rceil$) of the total solutions for Eq.~(\ref{eq22}), such eigenvalues belong to the original cycle and correspond to the eigenvalues whose first order corrections are $0$ (See Eq.~(\ref{eq29})) in the perturbation theory. In view of Fig.~\ref{fg1}, such eigenvalues are equivalent to the eigenvalues of a 1D chain of size $\lceil N/2\rceil$. The other eigenvalues are determined by $\Theta(N,m,x)=0$ (the second factor in the big bracket) in Eq.~(\ref{c9}).

Next, we prove $x_1=\pm x_m$ using Eq.~(\ref{c9}). We use Eq.~(\ref{eq18}) to write $x_{m-1}=-c_2x_m/c_1$, and substitute $x_{m-1}$ into Eq.~(\ref{c3}),
\begin{equation}\label{c10}
\begin{aligned}
\frac{x_1}{x_m}&= \bigg\{U_{m-2}(x)\frac{U_{{\scriptscriptstyle N}-m}(x)U_{m-3}(x)+U_{{\scriptscriptstyle N}-m}(x)+U_{{\scriptscriptstyle N}-2}(x)+1}{U_{{\scriptscriptstyle N}-1}(x)+U_{{\scriptscriptstyle N}-m}(x)U_{m-2}(x)}-U_{m-3}(x) \bigg\} \\
&=\frac{[U_{{\scriptscriptstyle N}-2}(x)U_{m-2}(x)-U_{{\scriptscriptstyle N}-1}(x)U_{m-3}(x)] +U_{{\scriptscriptstyle N}-m}(x)U_{m-2}(x)+ U_{m-2}(x) }{U_{{\scriptscriptstyle N}-1}(x)+U_{{\scriptscriptstyle N}-m}(x)U_{m-2}(x)} \\
&=\frac{U_{{\scriptscriptstyle N}-m}(x) +U_{{\scriptscriptstyle N}-m}(x)U_{m-2}(x)+ U_{m-2}(x) }{U_{{\scriptscriptstyle N}-1}(x)+U_{{\scriptscriptstyle N}-m}(x)U_{m-2}(x)}.\ \ \ \ \  U_{{\scriptscriptstyle N}-1}(x)+U_{{\scriptscriptstyle N}-m}(x)U_{m-2}(x)\neq0\ \ \ \ \ \ \ \ \ \ (\ref{a10})\ \ \text{used}
\end{aligned}
\end{equation}
If $T_{\scriptscriptstyle N}(x)-1=0$, according to Eq.~(\ref{eq22}), $U_{{\scriptscriptstyle N}-1}(x)=U_{{\scriptscriptstyle N}-m}(x)+U_{m-2}(x)$. Substitute it into Eq.~(\ref{c10}), we get $x_1/x_m=1$ (the denominator in Eq.~(\ref{c10}) not equal to $0$). If $\Theta(N,m,x)=0$ (the second factor in the big bracket in Eq.~(\ref{c9})), according to Eq.~(\ref{eq22}), $U_{{\scriptscriptstyle N}-m}(x)=U_{{\scriptscriptstyle N}-1}(x)+T_{\scriptscriptstyle N}(x)-U_{m-2}(x)-1$, thus $U_{{\scriptscriptstyle N}-m}(x)U_{m-2}(x)$ can be simplified as,
\begin{equation}\label{c11}
\begin{aligned}
U_{{\scriptscriptstyle N}-m}(x)U_{m-2}(x)&=[U_{{\scriptscriptstyle N}-1}(x)+T_{\scriptscriptstyle N}(x)-U_{m-2}(x)-1]U_{m-2}(x)
=-\frac{1}{2}[T_{\scriptscriptstyle N}(x)+2U_{{\scriptscriptstyle N}-1}(x)-1],
\end{aligned}
\end{equation}
where $\Theta(N,m,x)=0$ has been used in the above calculation. Substituting $U_{{\scriptscriptstyle N}-m}(x)+U_{m-2}(x)=U_{{\scriptscriptstyle N}-1}(x)+T_{\scriptscriptstyle N}(x)-1$ (Eq.~(\ref{eq22})) and $U_{{\scriptscriptstyle N}-m}(x)U_{m-2}(x)$ in (\ref{c11}) into Eq.~(\ref{c10}) lead to,
\begin{equation}\label{c12}
\begin{aligned}
\frac{x_1}{x_m}&=\frac{U_{{\scriptscriptstyle N}-1}(x)+T_{\scriptscriptstyle N}(x)-1-\frac{1}{2}[T_{\scriptscriptstyle N}(x)+2U_{{\scriptscriptstyle N}-1}(x)-1]}{U_{{\scriptscriptstyle N}-1}(x)-\frac{1}{2}[T_{\scriptscriptstyle N}(x)+2U_{{\scriptscriptstyle N}-1}(x)-1]}
=\frac{\frac{1}{2}[T_{\scriptscriptstyle N}(x)-1]}{\frac{1}{2}[1-T_{\scriptscriptstyle N}(x)]}=-1. \ \ \ \ T_{\scriptscriptstyle N}(x)\neq1
\end{aligned}
\end{equation}
Thus we have proved $x_1=\pm x_m$, where the sign depends on the eigenvalues. For the largest eigenvalue, $x_1=-x_m$ holds because the largest eigenvalue satisfies $\Theta(N,m,x)=0$ in Eq.~(\ref{c9}) ($T_{\scriptscriptstyle N}(x)\neq1$).

Now we prove that if $m$ (or $N+m$) is odd, the components $x_j$ at the symmetry axis $j=(m+1)/2$ of the eigenstates corresponding to eigenvalues satisfying $\Theta(N,m,x)=0$, {\em i.e.}, $x_j=0$ ($j=(m+1)/2$) for the eigenstates whose eigenvalues satisfy $\Theta(N,m,x)=0$ (See the second factor in the big bracket in Eq.~(\ref{c9})). According to Eq.~(\ref{eq12}), $x_j=U_{m-1-j}(x)x_{m-1}(x)-U_{m-2-j}(x)x_m$. Setting $j=(m+1)/2$ leads to $x_{(m+1)/2}=U_{(m-3)/2}(x)x_{m-1}(x)-U_{(m-5)/2}(x)x_m$. Considering $x_{m-1}=-c_2x_m/c_1$ (See $c_1$ and $c_2$ in Eq.~(\ref{b1}) and (\ref{b2})), we obtain,
\begin{equation}\label{c13}
\begin{aligned}
x_{j=(m+1)/2}&= \bigg\{U_{(m-3)/2}(x)\frac{U_{{\scriptscriptstyle N}-m}(x)U_{m-3}(x)+U_{{\scriptscriptstyle N}-m}(x)+U_{{\scriptscriptstyle N}-2}(x)+1}{U_{{\scriptscriptstyle N}-1}(x)+U_{{\scriptscriptstyle N}-m}(x)U_{m-2}(x)}-U_{(m-5)/2}(x) \bigg\}x_m \\
&=\bigg\{\frac{U_{{\scriptscriptstyle N}-m}(x)[U_{m-3}(x)U_{(m-3)/2}(x)-U_{m-2}(x)U_{(m-5)/2}(x)] + [U_{{\scriptscriptstyle N}-2}(x)U_{(m-3)/2}(x)-U_{{\scriptscriptstyle N}-1}(x)U_{(m-5)/2}(x)]}{U_{{\scriptscriptstyle N}-1}(x)+U_{{\scriptscriptstyle N}-m}(x)U_{m-2}(x)}\\
&\ \ \ \ \ +\frac{ U_{{\scriptscriptstyle N}-m}(x)U_{(m-3)/2}(x)+ U_{(m-3)/2}(x) }{U_{{\scriptscriptstyle N}-1}(x)+U_{{\scriptscriptstyle N}-m}(x)U_{m-2}(x)}\bigg\}x_m \ \ \ \ \ \ \ \ \ \ (\ref{a10})\ \ \text{can be used in the middle brackets}\\
&=\frac{2U_{{\scriptscriptstyle N}-m}(x)U_{(m-3)/2}(x) +U_{{\scriptscriptstyle N}-(m+1)/2}(x)+ U_{(m-3)/2}(x) }{U_{{\scriptscriptstyle N}-1}(x)+U_{{\scriptscriptstyle N}-m}(x)U_{m-2}(x)}x_m\equiv\frac{\Lambda x_m}{U_{{\scriptscriptstyle N}-1}(x)+U_{{\scriptscriptstyle N}-m}(x)U_{m-2}(x)}.
\end{aligned}
\end{equation}
The numerator $\Lambda$ in Eq.~(\ref{c13}) multiplied by $T_{(m-1)/2}(x)$ and applying the identity (\ref{a14}), we obtain $2T_{(m-1)/2}(x)U_{(m-3)/2}(x)=U_{m-2}(x)$ and $T_{(m-1)/2}(x)U_{{\scriptscriptstyle N}-(m+1)/2}(x)=\frac{1}{2}[U_{{\scriptscriptstyle N}-1}(x)+U_{{\scriptscriptstyle N}-m}(x)]$. Replacing the term $U_{{\scriptscriptstyle N}-m}(x)=U_{{\scriptscriptstyle N}-1}(x)+T_{\scriptscriptstyle N}(x)-U_{m-2}(x)-1$ (See Eq.~(\ref{eq22})) and after some appropriate algebraical calculations, we get,
\begin{equation}\label{c14}
\begin{aligned}
\Lambda T_{(m-1)/2}(x)=-U^2_{m-2}(x)+[T_{\scriptscriptstyle N}(x)+U_{{\scriptscriptstyle N}-1}(x)-1]U_{m-2}(x)+\frac{1}{2}[T_{\scriptscriptstyle N}(x)+2U_{{\scriptscriptstyle N}-1}(x)-1]\equiv-\frac{1}{2}\Theta(N,m,x).
\end{aligned}
\end{equation}
Compare Eq.~(\ref{c14}) and the second factor $\Theta(N,m,x)$ in Eq.~(\ref{c9}), the above equation equals to $0$ if the eigenvalues satisfying $\Theta(N,m,x) =0$. Thus $\Lambda=0$ and we have proved $x_j=0$ ($j=(m+1)/2$) for this case. Because the largest eigenvalue satisfies $\Theta(N,m,x)=0$, the component $x'_j$ at the symmetry axis $j=(m+1)/2$ equals to $0$ for the corresponding eigenstate. The proof for $j=(N+m+1)/2$ ($N+m$ is odd) is similar and we do not show here.
\section{Approximating the other eigenvalues using the determinant equation~(\ref{eq22})}
In this section, we approximate the other eigenvalues (except the largest eigenvalue) based on the determinant equation~(\ref{eq22}). Eq.~(\ref{c6}) can be rewritten as,
\begin{equation}\label{d1}
U_{m-2}(x)[T_{\scriptscriptstyle N}(x)-1]=U_{{\scriptscriptstyle N}-1}(x)T_{m-1}(x)-T_{\scriptscriptstyle N}(x)-U_{{\scriptscriptstyle N}-1}(x)+1
\end{equation}
Square both sides of the above equation and note that $U^2_{m-2}(x)=[T^2_{m-1}(x)-1]/(x^2-1)$ (Pell Equation in (\ref{a13})), we obtain,
\begin{equation}\label{d2}
\begin{split}
&\Big\{ (x^2-1)U^2_{{\scriptscriptstyle N}-1}(x)-[T_{\scriptscriptstyle N}(x)-1]^2 \Big\} T^2_{m-1}(x) +
 2\Big\{ (x^2-1)U^2_{{\scriptscriptstyle N}-1}(x)+(x^2-1)U_{{\scriptscriptstyle N}-1}(x)[T_{\scriptscriptstyle N}(x)-1] \Big\} T_{m-1}(x)\\
+&\Big\{ (x^2-1)U^2_{{\scriptscriptstyle N}-1}(x)+2(x^2-1)U_{{\scriptscriptstyle N}-1}(x)[T_{\scriptscriptstyle N}(x)-1]+x^2[T_{\scriptscriptstyle N}(x)-1]^2 \Big\}=0
\end{split}
\end{equation}
Replacing the term $U^2_{{\scriptscriptstyle N}-1}(x)$ by $U^2_{{\scriptscriptstyle N}-1}(x)=[T^2_{{\scriptscriptstyle N}-1}(x)-1]/(x^2-1)$ (See Pell Equation \ref{a13}) leads to,
\begin{eqnarray}\label{d3}
\begin{split}
[T_{\scriptscriptstyle N}(x)-1]\bigg\{ 2T^2_{m-1}(x)-2[T_{\scriptscriptstyle N}(x)+1+(x^2-1)U_{{\scriptscriptstyle N}-1}(x)]T_{m-1}(x) \\
+2(x^2-1)U_{{\scriptscriptstyle N}-1}(x)+x^2[T_{\scriptscriptstyle N}(x)-1]+ T_{\scriptscriptstyle N}(x)+1 \bigg\}=0
\end{split}
\end{eqnarray}
The first factor in the middle bracket $[\ ]$ in the left hand side $T_{\scriptscriptstyle N}(x^0)-1=0$ ($U_{{\scriptscriptstyle N}-1}(x^0)=0$) corresponds
to the eigenvalues for the regular cycle of $N$ nodes. Suppose the solutions of the second factor in the big bracket $\{\ \}$ being equal to $0$ can
be expanded as $x\approx x^0-\Delta$ ($x^0$ is the solution of $T_{\scriptscriptstyle N}(x^0)-1=0$, $\Delta$ is a small value compared to $x^0$),
the zero order approximation of $T_{m-1}(x)$, $T_{\scriptscriptstyle N}(x)$ and first order approximation of $(x^2-1)U_{{\scriptscriptstyle N}-1}(x)$ are given by
\begin{eqnarray}
&T_{m-1}(x)\approx T_{m-1}(x^0), \ \  T_{\scriptscriptstyle N}(x)\approx T_{\scriptscriptstyle N}(x^0)=1\\
&(x^2-1)U_{{\scriptscriptstyle N}-1}(x)\approx (x^0x^0-1)U_{{\scriptscriptstyle N}-1}(x^0)+
  x^0U_{{\scriptscriptstyle N}-1}(x^0)(x-x^0)+NT_{\scriptscriptstyle N}(x^0)(x-x^0)=-N\Delta.
\end{eqnarray}
Thus the second factor in the big bracket $\{\ \}$ of Eq.~(\ref{d3}) can be approximated as,
\begin{equation}\label{d5}
T^2_{m-1}(x^0)-(2-N\Delta)T_{m-1}(x^0)+(1-N\Delta)=0,
\end{equation}
which leads to solution $T_{m-1}(x^0)=1-N\Delta$ and the small deviation $\Delta=\frac{1}{N}[1-T_{m-1}(x^0)]$. Noting that $T_{m-1}(x^0)=\cos (m-1)\theta_n$ (the trigonometric function definition of the Chebyshev polynomials)~\cite{rn29,rn30}, our result obtained here accords with the first order correction in Eq.~(\ref{eq29}) by the perturbation theory.

\twocolumngrid

\end{document}